\documentclass[10pt,journal]{IEEEtran}
\usepackage{amsmath,amsfonts}
\usepackage{algorithm}
\usepackage{array}
\usepackage{subfigure}
\usepackage{textcomp}
\usepackage{stfloats}
\usepackage{url}
\usepackage{verbatim}
\usepackage{graphicx}
\usepackage{cite}
\usepackage{makecell}
\usepackage{qcircuit}
\usepackage{braket}
\usepackage{array}
\usepackage{slashed}
\usepackage{tikz}
\usepackage{stfloats}
\usepackage{makecell}
\usepackage{multirow}
\usepackage{enumitem}
\usepackage{url}
\usepackage{verbatim}
\usepackage{graphicx}
\usepackage{tabularx}
\usepackage{algpseudocode}
\usepackage{booktabs}

\newcommand{\circledsmall}[1]{\hbox{\tikz\draw (0pt, 0pt)
    circle (.45em) node {\makebox[0.15em][c]{\scriptsize#1}};}}
\newcommand{\circledtiny}[1]{\hbox{\tikz\draw (0pt, 0pt)
    circle (.3em) node {\makebox[0.15em][c]{\tiny#1}};}}

\begin{document}

\title{Attribute Fusion-based Evidential Classifier \\on Quantum Circuits}

\author{Hao~Luo, Qianli~Zhou$^{\ast}$, Lipeng~Pan, Zhen~Li, and~Yong~Deng$^{\ast}$
\thanks{The work is supported by National Natural Science Foundation of China (Grant No. 62373078)}
\thanks{Hao Luo is with the Institute of Fundamental and Frontier Science, University of Electronic Science and Technology of China, Chengdu 610054, China and also with Department of Communication Science and Engineering, School of Information Science and Technologies, Fudan University, Shanghai 200438, China.}
\thanks{Qianli Zhou (e-mail: zhouqianliuestc@hotmail.com)  and Lipeng Pan are with the Institute of Fundamental and Frontier Science, University of Electronic Science and Technology of China, Chengdu 610054, China}
\thanks{Zhen Li is with China Mobile Information Technology Center, Beijing 100029, China}
\thanks{Yong Deng is with the Institute of Fundamental and Frontier Science, University of Electronic Science and Technology of China, Chengdu 610054, China, and also with the School of Medicine, Vanderbilt University, Nashville 37240, USA. (e-mail: dengentropy@uestc.edu.cn)}
}

\markboth{IEEE TRANSACTIONS ON PATTERN ANALYSIS AND MACHINE INTELLIGENCE}%
{Shell \MakeLowercase{\textit{et al.}}: A Sample Article Using IEEEtran.cls for IEEE Journals}


\maketitle

\begin{abstract}
Dempster-Shafer Theory (DST) as an effective and robust framework for handling uncertain information is applied in decision-making and pattern classification. Unfortunately, its real-time application is limited by the exponential computational complexity. People attempt to address the issue by taking advantage of its mathematical consistency with quantum computing to implement DST operations on quantum circuits and realize speedup. However, the progress so far is still impractical for supporting large-scale DST applications. In this paper, we find that Boolean algebra as an essential mathematical tool bridges the definition of DST and quantum computing. Based on the discovery, we establish a flexible framework mapping any set-theoretically defined DST operations to corresponding quantum circuits for implementation. More critically, this new framework is not only uniform but also enables exponential acceleration for computation and is capable of handling complex applications. Focusing on tasks of classification, we based on a classical attribute fusion algorithm putting forward a quantum evidential classifier, where quantum mass functions for attributes are generated with a simple method and the proposed framework is applied for fusing the attribute evidence. Compared to previous methods, the proposed quantum classifier exponentially reduces the computational complexity to linear. Tests on real datasets validate the feasibility.
\end{abstract}


\section{Introduction}
\IEEEPARstart{D}{empster-Shafer} Theory (DST) \cite{dempsterProbabilities1967,shafer1976mathematical}, also referred to as evidence theory, is a mathematical framework to handle uncertain information. For a Frame of Discernment (FoD), DST assigns beliefs on the power set to describe the uncertain environment. Besides, evidence combination rules \cite{DENOEUX2008, Smet_alpha1997} enable DST to fuse the conflicted evidence from multiple sources  with different reliability levels. Thus, DST has been widely applied in decision-making  \cite{zhou2020assignment,decision_making_2_2002,fu2020evidential}, pattern classification \cite{Xiao2022NQMF,xiao2022generalized,xiao2022complex}, information fusion \cite{deng2023novel,information_fusion_1_2022,information_fusion_2_2013}. Besides, to handle uncertain complex data, DST framework is extended to the complex plane \cite{xiao2020generalize, Pan2022complex}. Since DST is defined on the power set with $2^n$ elements, the implementation of DST operations faces exponentially increasing computational complexity, which limits its real-time application and becomes a pressing problem of the moment. To solve the problem, some scholars have improved the classical algorithm \cite{barnettcompute1981, tessemComputation1993, wilsonMonteCarlo1991, benallaComputational2021}. However, their attempts do not provide a unified model for reducing computational complexity, as these algorithms require the pre-processing stage or are limited by assumptions on the input.

The above attempts fail as they are unable to break through the limitations of classical computers, where quantum computing might be an available option to bypass them. The natural parallel computing capability of quantum computers \cite{nielsen2002quantum} makes it an alternative to speed up classical algorithms which have been proven to exist in some specific application scenarios, such as Shor algorithm \cite{shor1999polynomial} for prime factorization and Grover search algorithm \cite{grover1996fast} for unstructured search problem. In the last decade, quantum computing has widely received attention \cite{Sheng2006switching, Ju2007quantum, Fyrigo2022Memristor, Laskar2023quantum}. In 2009,  Harrow \textit{et al.} proposed HHL algorithm \cite{harrow2009quantum} theoretically achieving exponential speedups in solving linear systems of equations on quantum circuits, which gives rise to many algorithms for Quantum Machine Learning (QML)  \cite{lloydQPCA2014, RebentrostQSVM2014,shi2023learning,tian2023learning}.  In addition, as the Noisy Intermediate-scale Quantum (NISQ) era \cite{preskillNISQ2018, huangComputing2023} will last for a long time, the study of Variational Quantum Algorithms (VQAs) \cite{cerezoVQA2021} currently becomes a hot direction. Using parameterized quantum circuits and classical parameter optimization, VQAs are applied to finding ground state \cite{peruzzoVQE2014,zeng2021vqe}, Hamiltonian diagonalization \cite{Zeng2021Hamiltonian}, searching quantum error-correcting codes \cite{zeng2022error}, mathematics \cite{vqls2019, vqaLinearAlgebra2021} under the current conditions of limited quantum resources. 

As to quantum attempts for DST, some scholars design quantum algorithms for implementation \cite{zhou2023bfqc, luo2023variational, panQuantum2022,he2023quantum,  xiao2023classify}, based on the connection between DST and the quantum model \cite{vourdasQuantum2014, Payandeh2021quantum}. Depending on the different points of interest in the mathematical consistency involved in both DST and quantum computing, current attempts can be divided into two categories. In the first category, it has been found that DST operation, essentially doing a matrix multiplication, can be modeled as a linear system and solved by quantum algorithms. The representative work is implementing matrix calculus on quantum circuits by HHL algorithm \cite{zhou2023bfqc}. Besides, to design an adapted option for the NISQ era, an idea is to solve the system of linear equations by VQAs \cite{luo2023variational}, with the help of the same tensor structure shared by belief matrices and multi-qubits quantum systems. However, the category of matrix-based quantum attempts cannot be applied in large-scale complex real-world applications such as clustering \cite{gong2021evidential} and classifiers \cite{liu2017combination, Hu2023attribute}, for the following reasons: (a) the computational complexity is still exponentially increasing since belief matrices are not sparse; (b) HHL algorithm is difficult to implement in the NISQ era \cite{Yalovetzky2021} and its arduous to use VQA where circuits deployment and iterative optimization require too much time \cite{luo2023variational}; (c) the conversion between quantum and classical information also takes exponential time complexity.

In contrast, the attempts falling in the second category are more straightforward, which focus on the correspondence between set-theoretic definitions of DST operations and controlled-NOT (CNOT) gates in quantum computing.  In \cite{zhou2023bfqc}, Zhou \textit{et al.} also corresponds one element to one qubit achieving the change of mass value for focal sets containing the element by manipulating the amplitude of the qubit. Further, they use CNOT gate to extract belief functions on quantum circuits.  Pan \textit{et al.} \cite{panQuantum2022} propose a quantum algorithm for implementing Dempster rule of combination and effectively reducing the complexity. Based on that, He \textit{et al.} \cite{he2023quantum} utilize the Toffoli gate to deploy the entire procedure of  Dempster combination rule, which is completely implemented on quantum circuits.  These methods well utilize the parallel capacity of quantum computing, using only a linear number of operations to manipulate exponential elements in the power set. However, the previous works are limited to a few DST operations, lacking a theoretically based unified framework for implementing more DST operations on quantum circuits.

In this paper, inspired by the work \cite{panQuantum2022}, we reconsider the original set-theoretic definition of the DST operation and its consistency with quantum computing. With the help of Boolean algebra, we link the set theory involved in DST operations with the logic involved in quantum gates, revealing the essential mathematical correspondence between both fields. On this basis, this paper presents a more flexible, efficient, and unified quantum framework for implementing any set-theoretically based DST operations such as negation and different combination rules. Applying the framework to the evidence combination in the classification problem, an attribute fusion-based evidential classifier is generated.

The structure of this paper is organized as follows. Section \ref{sec2} explains the primary definition of DST and quantum computing. Section \ref{sec3} introduces Boolean values into DST operations and quantum computing. The correspondence between focal sets and quantum states is established. Then, DST operations of negation and evidence combinations are presented in Boolean algebraic forms. A unified framework of quantum algorithms is proposed to implement the operations on quantum circuits. Simulations are performed to validate the quantum algorithm. In Section \ref{sec5}, based on the quantum algorithm, an attribute fusion-based evidential classifier on quantum circuits is proposed. The testing on real data sets demonstrates the feasibility and classification accuracy. Section \ref{sec6} summarizes the paper and discusses future research directions.

\section{\label{sec2}Preliminary}
\subsection{Dempster-Shafer Theory (DST)}
\subsubsection{Basic Definition}
For a problem to be handled, suppose that all the existing possible hypotheses $\theta^i$ form a set $\Theta$:
\begin{equation}\label{FoD}
    \Theta = \{ \theta^1,\theta^2,\dots,\theta^n \}.
\end{equation}
It is assumed that all the hypotheses are mutually exclusive and exhaustive. And $\Theta$ denotes an FoD of $n$ elements.

In the traditional Bayesian theory, probabilities are directly assigned to elements in $\Theta$. On the contrary, as an extension of the Bayesian theory, DST is defined on the power set, which is denoted as:
\begin{equation} \label{power_set}
    2^{\Theta}=\{\emptyset,\{\theta^{1}\},\dots,\{\theta^{n}\},\{\theta^{2}\theta^{1}\},\dots,\{\theta^{n}\dots\theta^{1}\}\}.
\end{equation}
And the mass function $m$ maps the beliefs to elements in $2^\Theta$. Here $m$ is also called basic belief assignment (BBA). The assigned beliefs represent the support degree for the propositions, which satisfies:
\begin{equation} \label{sum_mass}
    \sum_{F \in 2^\Theta} m(F) = 1.
\end{equation}
If $m(F)\neq 0$, $F$ is a focal set. 

\subsubsection{DST operations}
Dubois and Prade \cite{dubois1986} define the negation of a mass function as follows:
\begin{equation} \label{negation_def}
    \overline{m}(F) = m( \overline{F});
\end{equation}
where $\overline{F}$ is the complementary sets in set theory.

In addition, rules of combination are proposed to fuse mass functions generated by multiple sources. They are also defined in a set-theoretic way.  Conjunctive rule of combination (CRC) $\circledsmall{$\cap$}$ shown in Eq.(\ref{ccr_def_n})  and disjunctive rule of combination (DRC) $\circledsmall{$\cup$}$ shown in Eq.(\ref{dcr_def_n}) are two typical rules \cite{DENOEUX2008}. The set operation in the definition can be extended further \cite{Smet_alpha1997} to form the exclusive disjunctive rule $\circledsmall{$\underline{\cup}$}$ presented in Eq.(\ref{exclusive_disjunctive}) and any other customized rules e.g. taking Eq.(\ref{customized}) as an example in the paper. Suppose $m_r$ is the mass function to be fused, they are given by:
\begin{equation} \label{ccr_def_n}
\begin{aligned}
    m(F)=\sum_{F_1\cap \dots\cap F_p=F} \prod_{r=1}^p m_r(F_r), m=m_1\circledtiny{$\cap$}\dots\circledtiny{$\cap$}m_p;
\end{aligned}
\end{equation}
\begin{equation}\label{dcr_def_n}
\begin{aligned}
    m(F)=\sum_{F_1\cup \dots\cup F_p=F} \prod_{r=1}^p m_r(F_r),m=m_1\circledtiny{$\cup$}\dots\circledtiny{$\cup$}m_p;
\end{aligned}   
\end{equation}
\begin{equation}\label{exclusive_disjunctive}
    m(F)=\sum_{F_1\underline{\cup} F_2=F}m_1(F_1)m_2(F_2),m=m_1\circledsmall{$\underline{\cup}$}m_2;
\end{equation}
\begin{equation}\label{customized}
\begin{aligned}
    m(F)=\sum_{(\overline{F_1\cap F_2})\cap (F_2\cup F_3)=F}m_1(F_1)m_2(F_2)m_3(F_3),& 
    \\m=(\overline{m_1\circledsmall{${\cap}$}m_2})\circledsmall{${\cap}$}( m_2\circledsmall{${\cup}$}m_3).&
\end{aligned}  
\end{equation}
In Eq.(\ref{exclusive_disjunctive}), $\underline{\cup}$ denotes the symmetric difference in set theory, where $A\underline{\cup}B=(A\cap \overline{B})\cup (\overline{A}\cap B)$.

\subsection{Quantum Computing}
In this section, we briefly introduce basic concepts in quantum computing and quantum mass function that encodes mass values as the amplitudes of quantum states. For a more detailed discussion, please refer to Chapter 4 of the book "Quantum Computation and Quantum Information" \cite{nielsen2002quantum}.
\subsubsection{Qubits and Quantum Gates}

A qubit is a basic unit of information storage and processing in quantum computing. It is represented by the state $\ket{\psi}$ in a two-dimensional Hilbert space, expressed as $\ket{\psi}=\alpha\ket{0}+\beta\ket{1}$, where $\ket{0}=\begin{pmatrix} 1  \\ 0 \end{pmatrix}$ and $\ket{1}=\begin{pmatrix} 0  \\ 1 \end{pmatrix}$ are orthogonal ground states. The coefficients $\alpha$ and $\beta$ are complex numbers satisfying the normalization condition $|\alpha|^2+|\beta|^2=1$. When measured, the probability of obtaining $\ket{0}$ or $\ket{1}$ is $|\alpha|^2$ or $|\beta|^2$, respectively.

For multiple qubits $\ket{\psi_1},\dots,\ket{\psi_n}$, the overall state $\ket{\psi}$ is calculated using the tensor product $\otimes$ of each qubit's state. The resulting state is $\ket{\psi}=\ket{\psi_1}\otimes\dots\otimes\ket{\psi_n}$.

To simplify notation, a decimal representation $\ket{\cdot}_D$ is introduced for each ground state $\ket{0\dots00}$,$\ket{0\dots01}$,$\dots,\ket{1\dots11}$. Suppose $(i^ni^{n-1}\dots i^1)_2$ is the binary representation for a decimal number $i$, $\ket{i}_D$ is defined as $\ket{i^ni^{n-1}\dots i^1}=\ket{i^n}\otimes \ket{i^{n-1}}\otimes \dots \otimes \ket{i^1}$.

Some quantum gates encountered in this paper are presented in table \ref{tab:gate}.

\begin{table}[H]
\caption{\label{tab:gate}Quantum gates used in the paper.}
\centering
\begin{tabular}{c|c|c}
Gate & Matrix Representation& Circuit Symbol\\
\hline
X gate & $\begin{pmatrix}0 & 1 \\1 & 0\end{pmatrix}$& $\Qcircuit @C=1em @R=.7em {
& \gate{X} & \qw
}$\\
$R_Y$ gate &  $\begin{pmatrix}\cos(\alpha/2) & \sin(\alpha/2) \\-\sin(\alpha/2) & \cos(\alpha/2)\end{pmatrix}$ & $\Qcircuit @C=1em @R=.7em {
& \gate{R_\mathrm{Y}(\alpha)} & \qw
}$\\
\makecell{controlled-$X$  \\ (CNOT) gate} & $ \begin{pmatrix}1 & 0 & 0 & 0 \\0 & 1 & 0 & 0 \\ 0 & 0 & 0 & 1 \\0 & 0 & 1 & 0    \end{pmatrix}$ & $\Qcircuit @C=1em @R=.7em {
& \ctrl{1} &  \qw \\
& \targ &  \qw
}$\\
\makecell{controlled-$R_Y$\\ gate} &  $ \begin{pmatrix}1 & 0 & 0 & 0 \\0 & 1 & 0 & 0 \\ 0 & 0 & \cos(\alpha/2) & \sin(\alpha/2) \\0 & 0 & -\sin(\alpha/2) & \cos(\alpha/2)    \end{pmatrix}$ & $\Qcircuit @C=1em @R=.7em {
& \ctrl{1} & \qw \\
& \gate{R_Y(\alpha)} & \qw }$\\
\end{tabular}
\end{table}

\subsubsection{Quantum Mass Function}
Since $n$ qubits generate $2^n$ mutually orthogonal superposition, the $2^n$ mass values in the power set can be encoded on their amplitudes.  Let the square root value of the mass for the $i$-th focal set be encoded as the amplitude for the $i$-th ground state $\ket{i}_D$ and obtain the corresponding quantum mass function $\ket{m}$. 
\begin{equation}\label{qmf}
    \ket{m}=\sum_{i=0}^{2^n-1}\sqrt{m(\mathcal{F}(i))}\ket{i}_D.
\end{equation}
Thus the measurement probability of the ground state $\ket{i}_D$ exactly equals the mass value, whose sum is just enough to satisfy the quantum normalization condition according to Eq.(\ref{sum_mass}).

\begin{figure}[H]
\centering
\includegraphics[width=8.6cm]{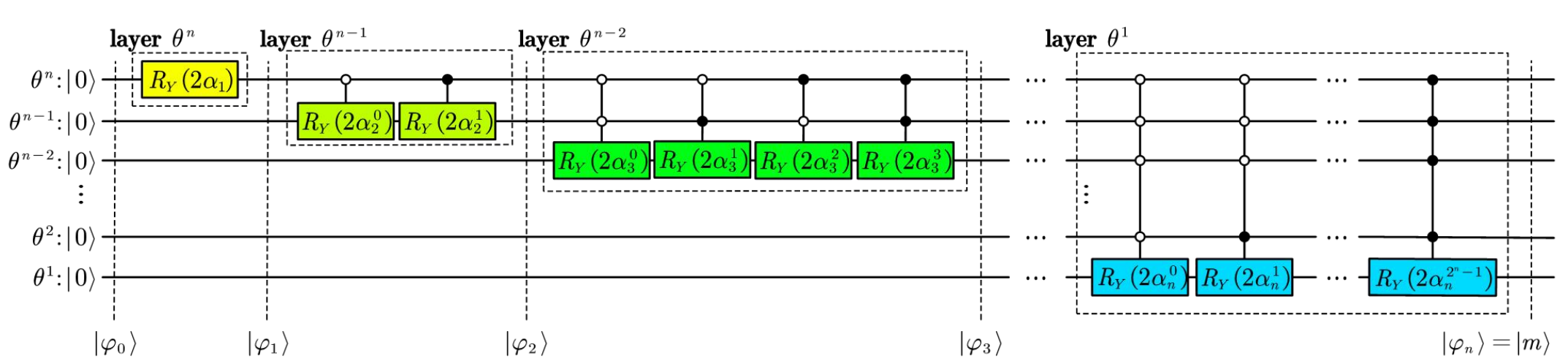}
\caption{\label{diag_tree}The tree-like memory structure of the quantum circuits for preparing a quantum mass function.}
\end{figure}

For any  mass functions, Zhou \textit{et al.} \cite{zhou2023bfqc} has already proposed a $R_Y$ gate implementation method with $n$ layers, which is called a tree-like memory structure as presented in Fig.\ref{diag_tree}.

After $n$ layers, the initial state $\ket{0}^{\otimes n}$ transforms to  $\ket{m}$ in Eq.(\ref{qmf}). However, the method has the disadvantage of its exponential computational complexity $\mathcal{O}(n2^n)$, which requires further speedup for real application scenarios. 

\section{\label{sec3}Operate Mass Function on Quantum Circuits}

\subsection{Introduce Boolean Values to DST and Quantum States}
 From a new perspective of Boolean values, we reconsider the representation for  focal sets in DST and quantum ground states. Further,  the correspondence between DST and quantum computing is established through Boolean values.

\subsubsection{Boolean Values for Focal Set}
Under an FoD of $n$ elements defined in Eq.(\ref{FoD}), suppose $F$ is a focal set in the power set. $F$ can be determined by a series of Boolean values $i^k,k=1,\dots,n$, where $i^k$ represents whether the element $\theta^k$ belongs to $F$. If it is true then $i^k$ is 1, otherwise it is 0.
\begin{equation}\label{ik}
    i^k=\left\{\begin{array}{l}
        1,\quad\theta^k\in F \\
        0,\quad\theta^k \notin F
\end{array}\right.,\quad k=1,\dots,n.
\end{equation}
$F$ can be uniquely determined by these Boolean values $i^n,\dots,i^1$. For convenient representation in decimal, the index number $i$ is introduced.
\begin{equation}\label{i}
    i=(i^ni^{n-1}\dots i^1)_2=2^{n-1}i^n+2^{n-2}i^{n-1}+\dots+2^0i^1.
\end{equation}
Since $i^n,\dots,i^1$ are all Boolean values, the range of $i$ is from 0 to $2^n-1$. Therefore, for any element in the power set, unique corresponding  Boolean values can be found along with a corresponding decimal index. So a mapping $\mathcal{F}$ can be created between the index and the elements in the power set.
\begin{equation}\label{index_powerset}
   \mathcal{F}:i=(i^ni^{n-1}\dots i^1)_2\in \mathbb{Z} \rightarrow 2^\Theta.
\end{equation}

For the case of $n=3$, the correspondence is shown in the following table \ref{index_table}.

\subsubsection{Boolean Values for Quantum States}
For a quantum system of $n$ qubits, it generates $2^n$ mutually orthogonal quantum ground states. They can be represented by $n$ Boolean values $\ket{i^ni^{n-1}\dots i^1}$, whose decimal representation is $\ket{i}_D$. Since Boolean values are mapped to focal sets, each quantum ground state corresponds to each focal set through the representation of Boolean values, which is presented in table \ref{index_table}.

\begin{table}[!t]
\centering
\caption{\label{index_table}%
Correspondence between indexes, Boolean values, and focal sets under an FoD of 3 elements.}
\begin{tabular}{c|c|c|c|c}
Index: $i$ & 0&1&2&3\\
Boolean value: $i^3i^2i^1$ & 000&001&010&011\\
Focal Set: $\mathcal{F}(i^3i^2i^1)$ & $\emptyset$ & $\{\theta^1\}$  & $\{\theta^2\}$&$\{\theta^2\theta^1\}$\\
Quantum State $\ket{i^3i^2i^1}$& $\ket{000}$ & $\ket{001}$ & $\ket{010}$  & $\ket{011}$\\
\hline
Index: $i$ & 4&5&6&7\\
Boolean value: $i^3i^2i^1$ & 100&101&110&111\\
Focal set: $\mathcal{F}(i^3i^2i^1)$ & $\theta^3$ & $\{\theta^3\theta^1\}$  & $\{\theta^3\theta^2\}$&$\{\theta^3\theta^2\theta^1\}$\\
Quantum State $\ket{i^3i^2i^1}$& $\ket{100}$ & $\ket{101}$ & $\ket{110}$  & $\ket{111}$\\
\end{tabular}
\end{table}

Thus the quantum mass function $\ket{m}$ defined in Eq.(\ref{qmf}) is equivalent to the following Boolean form:

\begin{equation} \label{ideal_state}
    \begin{aligned}
        \ket{m} 
        =\sum_{i^n=0}^1\dots\sum_{i^1=0}^1\sqrt{m(\mathcal{F}(i^n\dots i^1))}\ket{i^n\dots i^1}
    \end{aligned}
\end{equation}

\subsection{Implement DST Operations on Quantum Circuits}
Since Boolean values has been introduced for focal sets,  the Boolean algebraic form of DST operations can be derived including the negation of mass function and multiple rules of combination. Next, we hope to utilize quantum circuits to implement Boolean algebra to achieve DST operations. 

\subsubsection{Boolean Algebra for DST Operations}
The set operators involved in the definitions of DST operations, including complement $\bar{\quad}$, intersection $\cap$, and union $\cup$ in set theory, can be represented by Boolean algebra.

First, for the negation, the complement is involved in its definition in Eq.(\ref{negation_def}). Suggest that $F_1$ and $F_2$ are complementary sets of each other. For a single element $\theta^k$ in FoD, it either belongs to $F_1$ or its complement set $F_2$, which can derive that
\begin{equation} \label{complement_2}
    F_2=\overline{F_1}\quad \Rightarrow \quad i_2^k= \neg i_1^k;
\end{equation}
where $i_r$ and $i_r^k$ denote the index and Boolean values for $F_r$, respectively. Therefore, the definition of the negation in Eq.(\ref{negation_def}) can be expressed with Boolean algebra:
\begin{equation}\label{negation_Boolean}
    \overline{m}(\mathcal{F}(i^ni^{n-1}\dots i^1)) = m(\mathcal{F}(\neg i^n \neg i^{n-1}\dots \neg i^1)).
\end{equation}

Besides, the intersection and union in set theory can be transformed into the form of Boolean algebra with AND operator $\land$ and OR operator $\vee$.
\begin{equation} \label{intersection}
\begin{aligned}
     F = F_1 \cap F_2\cap \dots \cap F_p \quad \Rightarrow \quad i^k=i_1^k \land i^k_2\land \dots \land i^k_p;
\end{aligned}
\end{equation}
\begin{equation} \label{union}
\begin{aligned}
        F = F_1 \cup F_2\cup \dots \cup F_p \quad \Rightarrow \quad i^k=i_1^k \vee i^k_2\vee \dots \vee i^k_p.
\end{aligned}
\end{equation}
Similarly, since CRC and DRC respectively utilize the intersection and union operations, one can obtain the variation of the definitions Eq.(\ref{ccr_def_n}) and (\ref{dcr_def_n}) with Boolean algebra.

\begin{equation}\label{ccr_Boolean}
\begin{aligned}
        m(\mathcal{F}(i^n\dots i^1))&= \sum_{i_1^n \land \dots\land i_p^n=i^n} \dots  \sum_{i_1^1 \land \dots\land i_p^1=i^1}  \\&\prod_{r=1}^p m_r(\mathcal{F}(i_r^n\dots i_r^1)),\quad m=m_1\circledtiny{$\cap$}\dots\circledtiny{$\cap$}m_p;
\end{aligned}
\end{equation}
\begin{equation}\label{dcr_Boolean}
\begin{aligned}
    m(\mathcal{F}(i^n\dots i^1))&= \sum_{i_1^n \vee \dots\vee i_p^n=i^n} \dots  \sum_{i_1^1 \vee \dots\vee i_p^1=i^1}  \\& \prod_{r=1}^p m_r(\mathcal{F}(i_r^n\dots i_r^1)),\quad m=m_1\circledtiny{$\cup$}\dots\circledtiny{$\cup$}m_p.
\end{aligned}
\end{equation}

Compared to the original definition, the involved set operations $F_1\cap \dots \cap F_p=F$ and $F_1\cup \dots \cup F_p=F$ are replaced by $n$ summation operations that are bounded by the Boolean algebra. One can extend the basic Boolean operators (NOT, AND, OR) to derive Boolean algebraic forms of more complex set-theoretic definitions e.g. the exclusive disjunctive rule in Eq.(\ref{exclusive_disjunctive}) and the customized rule in Eq.(\ref{customized}).
    \begin{equation}\label{exclusive_disjunctive_Boolean}
\begin{aligned}
    m(\mathcal{F}(i^n\dots i^1))
    =\sum_{(i_1^n \land \neg i_2^n )\vee(\neg i_1^n \land i_2^n)=i^n} \dots\sum_{(i_1^1 \land \neg i_2^1 )\vee(\neg i_1^1 \land i_2^1)=i^1} &\\ \prod_{j=1}^2 m_r(\mathcal{F}(i_r^n\dots i_r^1)),\quad m=m_1\circledsmall{$\underline{\cup}$}m_2&;
\end{aligned}
\end{equation}
\begin{equation}\label{customized_Boolean}
\begin{aligned}
m(\mathcal{F}(i^n\dots i^1))
    = \sum_{(\neg(i_1^n \land i_2^n))\land(i_2^n\vee i_3^n)=i^n} \dots \sum_{(\neg(i_1^1 \land i_2^1))\land(i_2^1\vee i_3^1)=i^1} &\\  \prod_{r=1}^3 m_r(\mathcal{F}(i_r^n\dots i_r^1)),
    \quad m=(\overline{m_1\circledsmall{${\cap}$}m_2})\circledsmall{${\cap}$}( m_2\circledsmall{${\cup}$}m_3)&.
\end{aligned}
\end{equation}

\subsubsection{Negation of a Mass Function on Quantum Circuits}
\begin{figure}[H]
\centering
\includegraphics[width=5cm]{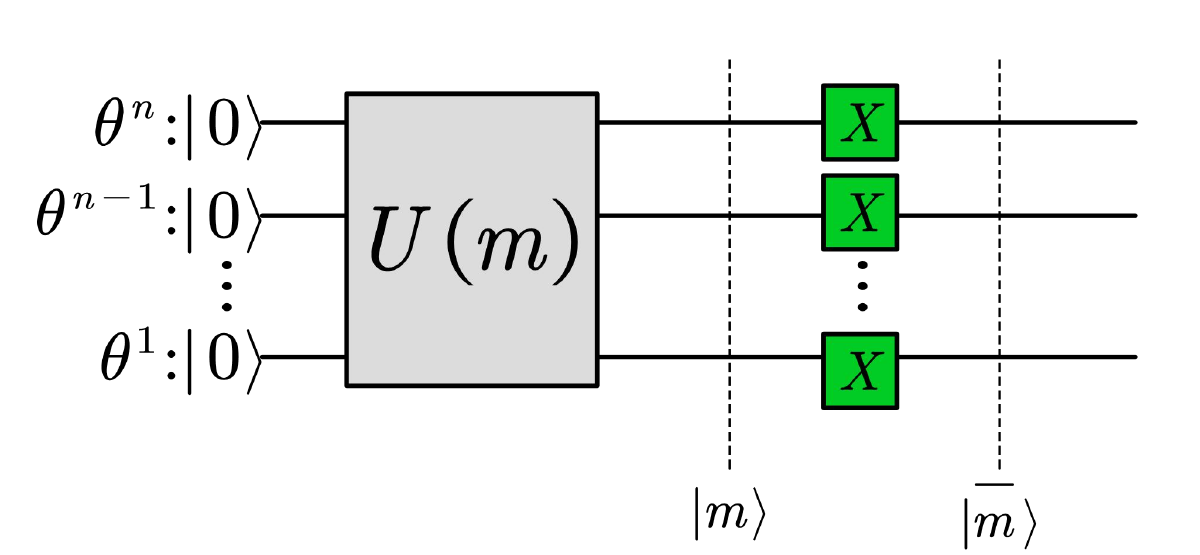}
\caption{\label{diag_negation} The quantum circuit for negation. }
\end{figure}

The Boolean algebraic form of the negation of a mass function is expressed in Eq.(\ref{negation_Boolean}), where each Boolean value should satisfy NOT operation in Eq.(\ref{complement_2}). Since the Pauli-$X$ gate is the quantum equivalent of  the classical NOT gate, mapping $\ket{0}$ to $\ket{1}$ and $\ket{1}$ to $\ket{0}$,  Eq.(\ref{complement_2}) can be realized by applying one $X$ gate on the corresponding qubit to convert $\ket{i^k}$ to $\ket{\neg i^k}$. Thus with the quantum mass function, the quantum version of the negation in Eq.(\ref{negation_Boolean}) can be represented on quantum circuits by applying $n$ Pauli-$X$ gates on each qubit after preparing $\ket{m}$. The quantum circuit is shown in Fig.\ref{diag_negation}.

After $X$ gates, the original quantum mass function $\ket{m}$ in Eq.(\ref{ideal_state}) becomes the state $\ket{\overline{m}}$, which represents the quantum mass function of $\overline{m}$. The proof is as follows.
\begin{equation}\label{prove_negation}
\begin{aligned}
    \ket{\psi}
        &=\sum_{i^n=0}^1\dots\sum_{i^1=0}^1\sqrt{m(\mathcal{F}(i^n\dots i^1))}\ket{\neg i^n \dots \neg i^1}\\
        &=\sum_{i^n=0}^1\dots\sum_{i^1=0}^1\sqrt{m(\mathcal{F}(\neg i^n \dots \neg i^1))}\ket{i^n\dots i^1}=\ket{\overline{m}}.
\end{aligned}
\end{equation}

\subsubsection{CRC on Quantum Circuits}
\begin{figure}[H]
\centering
\subfigure[CRC]{\includegraphics[width=.47\columnwidth]{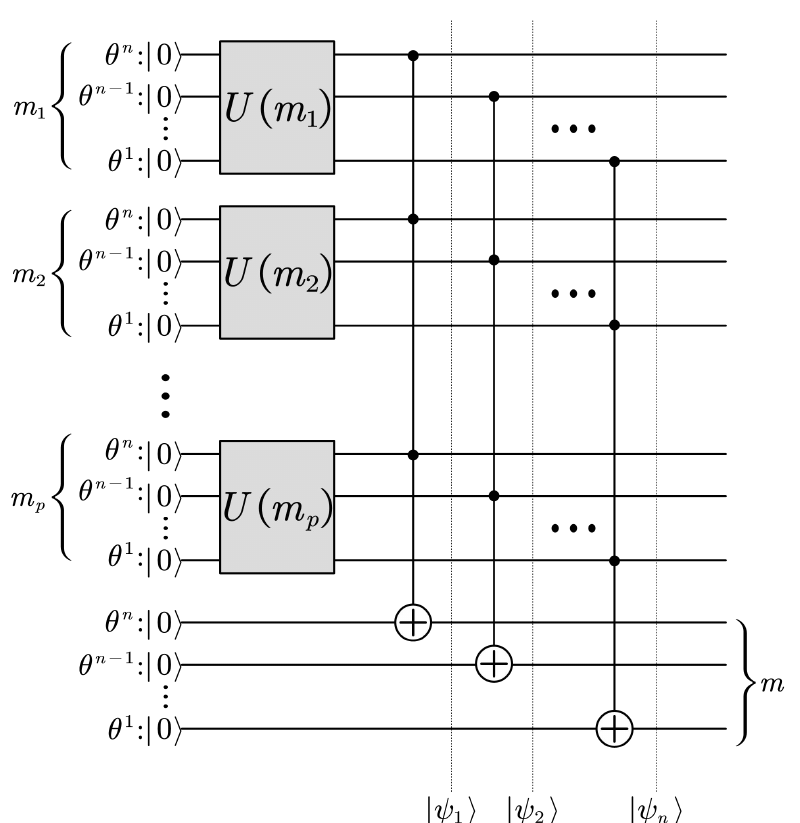}
\label{diag_ccr}}
\hfill
\subfigure[DRC]{\includegraphics[width=.47\columnwidth]{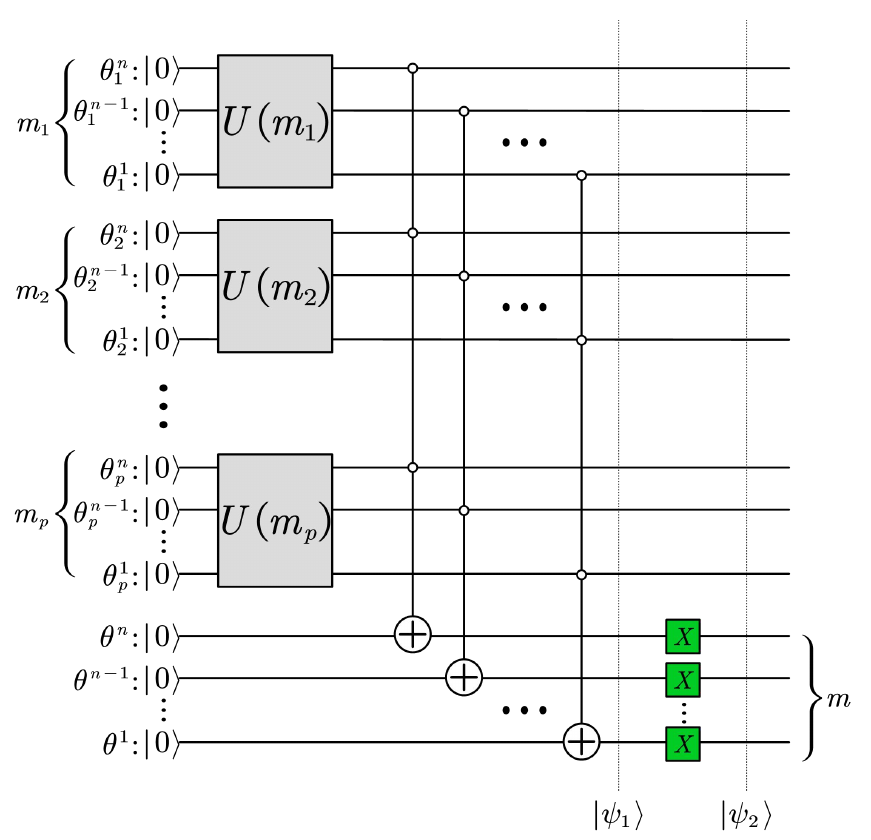}
\label{diag_dcr}}
\\
\subfigure[Exclusive disjunctive]{\includegraphics[width=.47\columnwidth]{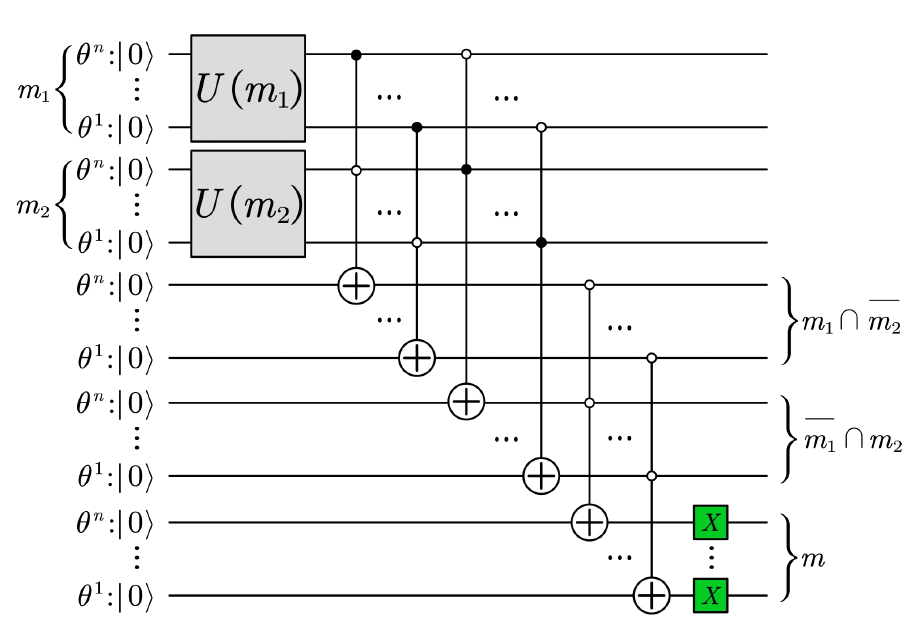}
\label{diag_exclusive}}
\hfill
\subfigure[Customized]{\includegraphics[width=.47\columnwidth]{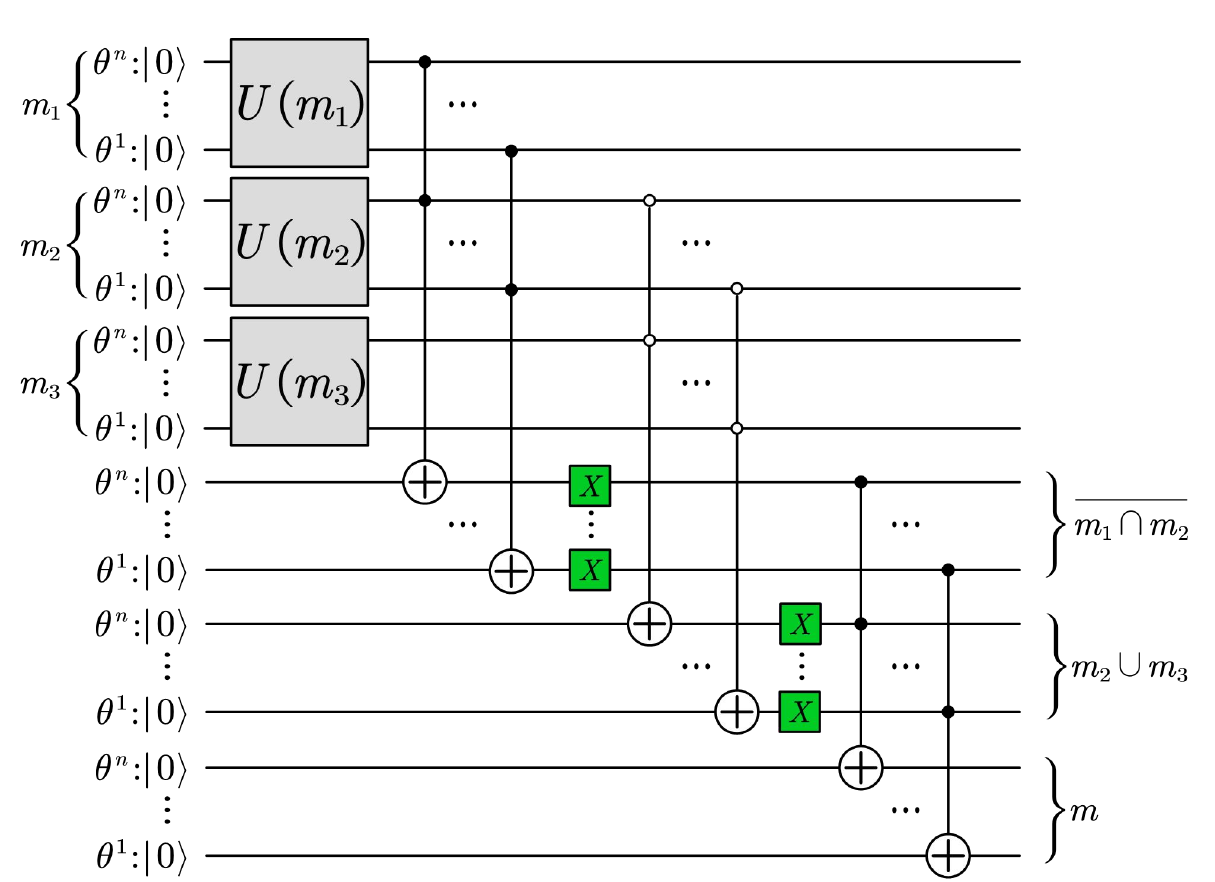}
\label{diag_custom}}
\caption{Quantum circuits for multiple rules of combination.}
\end{figure}

Consistent with the case of negation, to achieve the Boolean algebraic form of CRC in Eq.(\ref{ccr_Boolean}), Boolean values are required to satisfy the restriction of conjunction in Eq.(\ref{intersection}). The conjunction suggests that the result is 1 only if all Boolean values are 1, otherwise, it is 0. Actually, in quantum mass functions, one can pick one of the superpositions with all $\ket{1}$ by using the CNOT gate and setting all the qubits associated with these Boolean values as the control qubits. In this case, the target qubit flips only if control qubits are all in the state $\ket{1}$. Based on the idea, the quantum circuit displayed in Fig.\ref{diag_ccr} is designed to realize CRC for $p$ different mass functions.

The first step is preparing the initial quantum mass function for $m_1,\dots,m_p$ respectively. Second, the conjunction in Eq.(\ref{intersection}) is realized by utilizing CNOT gate element by element from $\theta^n$ to $\theta^1$. After $n$ CNOT operations, the final state can be expressed in Eq.(\ref{ccr_psin}). The state is be mathematically proven in Appendix 1.

    \begin{equation} \label{ccr_psin}
    \begin{aligned}
             \ket{\psi_n}&=\sum_{j^n=0}^1 \dots \sum_{j^{1}=0}^1 \Bigg\{\sum_{i_1^n\wedge\dots\wedge i_p^n=j^n} \dots \sum_{i_1^{1}\wedge\dots\wedge i_p^{1}=j^{1}} \\&\prod_{r=1}^{p}\sqrt{m_r(\mathcal{F}(i_r^n  \dots i_r^1))} \bigotimes_{r=1}^{p}\ket{i_r^n  \dots i_r^1} \ket{j^n\dots j^{1}}\Bigg\}.   
    \end{aligned}
    \end{equation}

Due to the orthogonality of ground states, the amplitude $A$ of the superposition $\ket{j^n\dots j^1}$  for the last $n$ qubits can be extracted as displayed below. 

\begin{equation}\label{ccr_amplitude}
\begin{aligned}
    &A(\ket{j^nj^{n-1}\dots j^1})\\&=\sqrt{\sum_{i_1^n\wedge\dots\wedge i_p^n=j^n}  \dots \sum_{i_1^{1}\wedge\dots\wedge i_p^{1}=j^{1}}\prod_{r=1}^{p} m_r(\mathcal{F}(i_r^n \dots i_r^1))  }\\
    &=\sqrt{m(\mathcal{F}(j^n\dots j^1))},\quad m=m_1\circledtiny{$\cap$}\dots\circledtiny{$\cap$}m_p.
\end{aligned}
\end{equation}

And the amplitude exactly corresponds to the square root value of the combined mass function defined in Eq.(\ref{ccr_Boolean}). Thus, in terms of the amplitude, the final state can be the quantum mass function after combining by CRC. Compared to the quantum mass function constructed by the tree-like memory structure in Section 3.2, the only difference is that this state is entangled with other parts in quantum circuits. 

In order to convert it to the classical version, one can measure the last $n$ qubits and record the probability $P(j^nj^{n-1}\dots j^1)$, which is equal to the mass after CRC.

\begin{equation}\label{ccr_measure}
\begin{aligned}
    &P(j^nj^{n-1}\dots j^1)=|A(\ket{j^nj^{n-1}\dots j^1})|^2
    \\&=m(\mathcal{F}(j^nj^{n-1}\dots j^1)),\quad m=m_1\circledtiny{$\cap$}\dots\circledtiny{$\cap$}m_p.
\end{aligned}
\end{equation}

\subsubsection{DRC on Quantum Circuits}
Similar to the conjunctive rule, one has to represent the disjunction in Eq.(\ref{union}) on quantum circuits to achieve the Boolean algebraic form of DRC expressed in Eq.(\ref{dcr_Boolean}). Since disjunction cannot be obviously represented by quantum gates whereas quantum circuits have been proposed to realize negation and conjunction in previous sections, we consider utilizing De Morgan's laws to replace the disjunctive regulation in Eq.(\ref{union}) by combining negation and conjunction.
\begin{equation}\label{disjunctive_De}
\begin{aligned}
    i^k=i_1^k \vee i^k_2\vee \dots \vee i^k_p  \Rightarrow  i^k=\neg(\neg i_1^k \wedge \neg i^k_2\wedge \dots \wedge \neg i^k_p)
\end{aligned} 
\end{equation}

According to Eq.(\ref{disjunctive_De}), DRC can be implemented in two steps. First, use CNOT gates to achieve $\neg i_1^k \wedge \neg i^k_2\wedge \dots \wedge \neg i^k_p$, which utilize similar quantum circuits for CRC. But in this case, the target qubit is set to flip when all control qubits are $\ket{1}$ instead of $\ket{0}$ because NOT applies before AND for each Boolean value. The control position is represented by a hollow point in the diagram. Second, achieve the NOT operations by implementing $X$ gates, with the same quantum circuits for negation. Therefore, by combining the two steps, the quantum circuits for DRC are displayed in Fig.\ref{diag_dcr}.

According to the proof displayed in Appendix 2, the final state $\ket{\psi_2}$ is:

\begin{equation}\label{dcr_final}
    \begin{aligned}
        \ket{\psi_2}&=\sum_{j^n=0}^1 \dots \sum_{j^{1}=0}^1 \Bigg\{\sum_{ i_1^n\vee\dots\vee  i_p^n= j^n} \dots \sum_{ i_1^{1}\vee\dots\vee  i_p^{1}= j^{1}} \\&\quad \quad\prod_{r=1}^{p}\sqrt{m_r(\mathcal{F}(i_r^n  \dots i_r^1))} \bigotimes_{r=1}^{p}\ket{i_r^n  \dots i_r^1}\ket{j^n\dots j^{1}}\Bigg\}.
    \end{aligned}
\end{equation}

Similar to Eq.(\ref{ccr_amplitude})(\ref{ccr_measure}), the final $n$ qubits contain the quantum mass function after DRC, which can be directly obtained by measurement probability.
\begin{equation}\label{dcr_probability}
\begin{aligned}
            &P(j^nj^{n-1}\dots j^1)=|A(\ket{j^nj^{n-1}\dots j^1})|^2
    \\&=m(\mathcal{F}(j^nj^{n-1}\dots j^1)),\quad m=m_1\circledtiny{$\cup$}\dots\circledtiny{$\cup$}m_p.
\end{aligned}
\end{equation}

\subsubsection{Extend to Other General Rules}
In fact, the construction mentioned above of quantum circuits can be extended to realize any rules of combination derived from set-theoretic definitions. The steps can be summarized in the following three steps.
\begin{itemize}
    \item \textbf{Step 1: }Transform the original definition of rules of combination into a Boolean algebraic version based on Boolean values for focal sets.
    \item \textbf{Step 2: }Analyze the Boolean values for each element of the FoD individually to obtain the constraints of Boolean algebra, which can be represented by the combination of three basic operators NOT, AND, OR.
    \item \textbf{Step 3: }For the derived Boolean algebraic constraints, decompose it into multiple stages, each of which involves only one basic Boolean operator. The circuit structure construction method for implementing the basic Boolean operators is  displayed below.
    \begin{itemize}
        \item NOT $\neg$: Use $X$ gate to flip every qubit.
        \item AND $\land$: Group the qubits according to their corresponding elements, and operate each group by CNOT gate. Set the qubits in the group as control qubits, and introduce a new qubit as the target qubit. If a Boolean value is operated by NOT operation before the AND operation, the flip is applied when the corresponding control qubit is $\ket{0}$ instead of $\ket{1}$.
        \item OR $\vee$: Utilize De Morgan's laws to convert OR operation into the operation of NOT, AND that can be implemented on quantum circuits.
    \end{itemize}
\end{itemize}

According to the proposed method, one can construct the quantum circuits for every stage and combine all the stages to form the  final quantum circuits.

Below, we give specific examples of implementing exclusive disjunction and the customized rules of combination to explore the process to transform any of the set-theoretically based rules of combination into a quantum circuit implementation by the proposed three-step approach.

The exclusive disjunction rule of combination is originally defined in Eq.(\ref{exclusive_disjunctive}). In Step 1, the Boolean algebraic form has been obtained and expressed in Eq.(\ref{exclusive_disjunctive_Boolean}). Then the corresponding Boolean constraints in Step 2 can be derived:
\begin{equation} \label{exclusive_disjunctive_element}
   (i_1^k \land \neg i_2^k )\vee(\neg i_1^k \land i_2^k)=i^k,\quad k=1,\dots,n.
\end{equation}

In Step 3, according to the logical order, the above constraints can be divided into three stages: (1) AND: get $i_1^k \land \neg i_2^k$; (2) AND: get $\neg i_1^k \land i_2^k$; (3) OR: finally get $(i_1^k \land \neg i_2^k )\vee(\neg i_1^k \land i_2^k)$. Since each stage only contains one basic Boolean operation, one can construct the quantum circuits stage by stage following the method in Step 3. The final quantum circuit structure for implementing the exclusive disjunctive rule of combination is shown in Fig.\ref{diag_exclusive}.

For the rule of combination customized by our own, in Step 1 the set-theoretical definition in Eq.(\ref{customized}) can be transformed into the Boolean algebraic version expressed in Eq.(\ref{customized_Boolean}). In Step 2, by analyzing each element in FoD alone, a series of constraints on the Boolean algebra is generated as below.
\begin{equation} \label{customized_element}
    (\neg(i_1^k \land i_2^k))\land(i_2^k\vee i_3^k)=i^k,\quad k=1,\dots,n.
\end{equation}

In Step 3,  quantum circuits can be constructed by dividing the constraint into the following implementation stages: (1) AND: get $i_1^k \land i_2^k$; (2) NOT: get $\neg(i_1^k \land i_2^k)$; (3) OR: get $i_2^k\vee i_3^k$; (4) AND: finally get $(\neg(i_1^k \land i_2^k))\land(i_2^k\vee i_3^k)$. Each implementation stage only involves one basic Boolean operator that corresponds to the structure of the circuit. Combining these stages of the circuit creates the quantum circuits for implementing the customized rule of combination, which is shown in Fig.\ref{diag_custom}.

\subsection{\label{sec4}Simulation}
To prove that the various rules of combination can be implemented on quantum circuits with the proposed method, we simulate CRC, DRC, the exclusive disjunctive, and also the customized rules of combination in this section. Simulation is conducted on the Qiskit platform.

\begin{figure}[!t]
\centering
\begin{minipage}{0.48\linewidth}
  \centerline{\includegraphics[width=3.5cm]{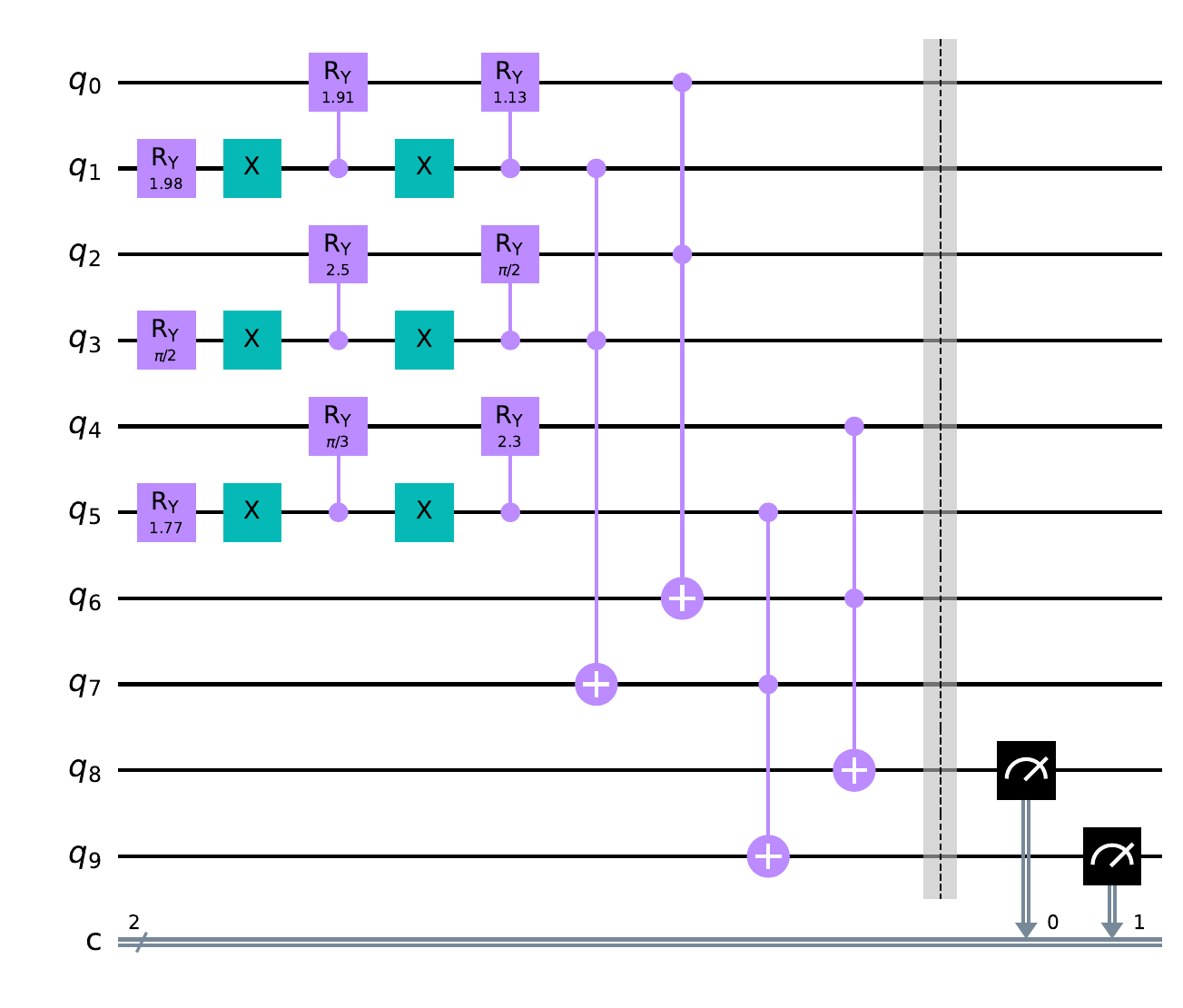}}
  \centerline{(a) Quantum circuit}
\end{minipage}
\hfill
\begin{minipage}{.48\linewidth}
  \centerline{\includegraphics[width=3.5cm]{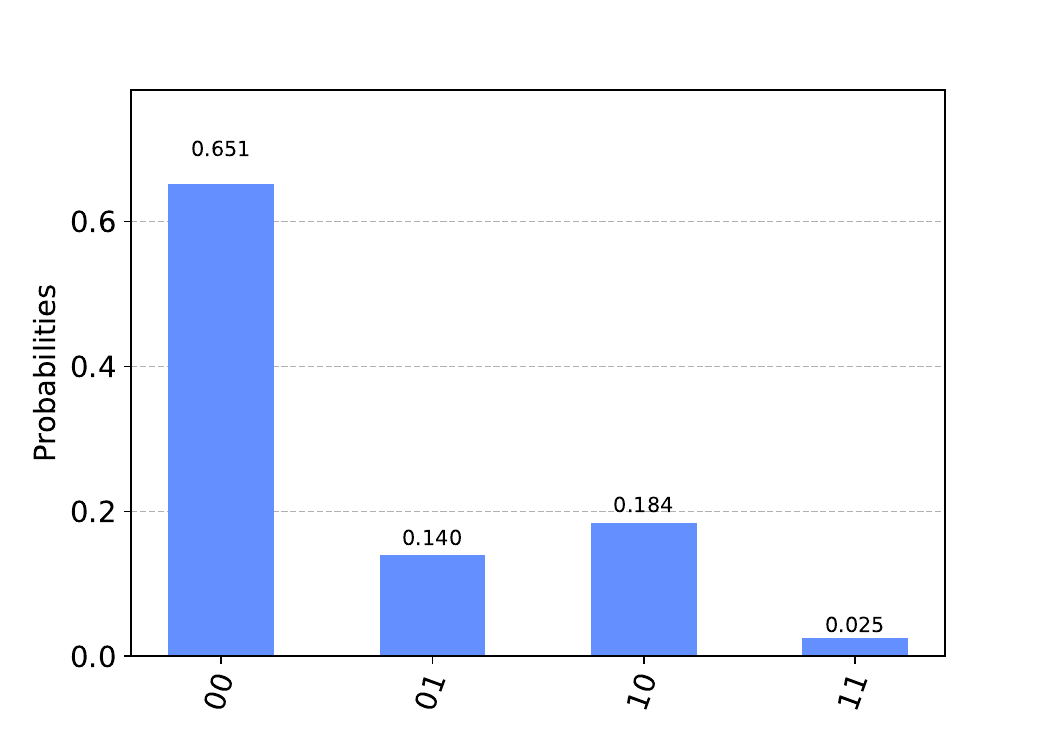}}
  \centerline{(b) Probability results}
\end{minipage}
\caption{Simulation for CRC. }
\label{ccr_simulate}
\end{figure}

\begin{figure}[!t]
\centering
\begin{minipage}{0.48\linewidth}
  \centerline{\includegraphics[width=3.5cm]{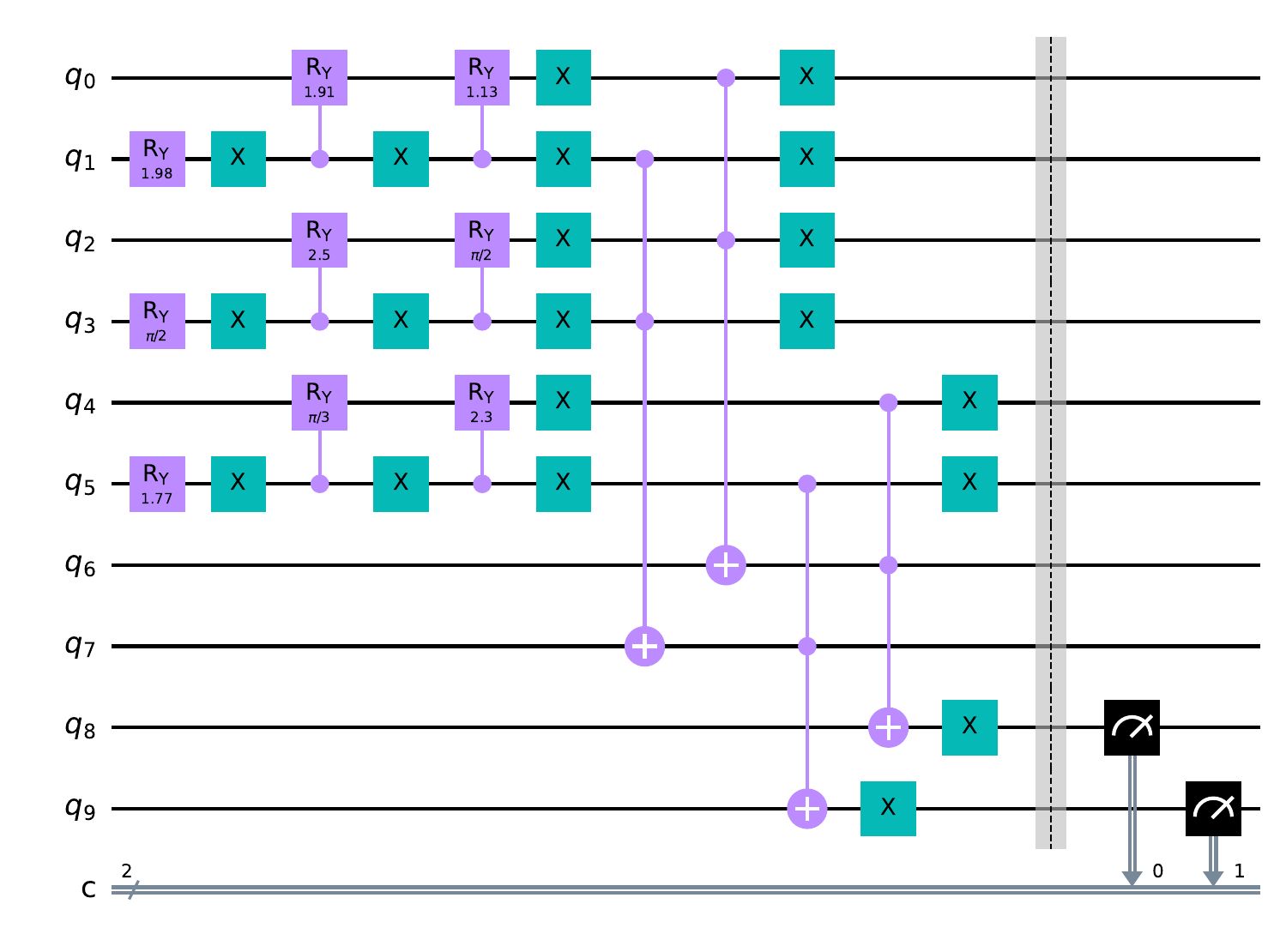}}
  \centerline{(a) Quantum circuit}
\end{minipage}
\hfill
\begin{minipage}{.48\linewidth}
  \centerline{\includegraphics[width=3.5cm]{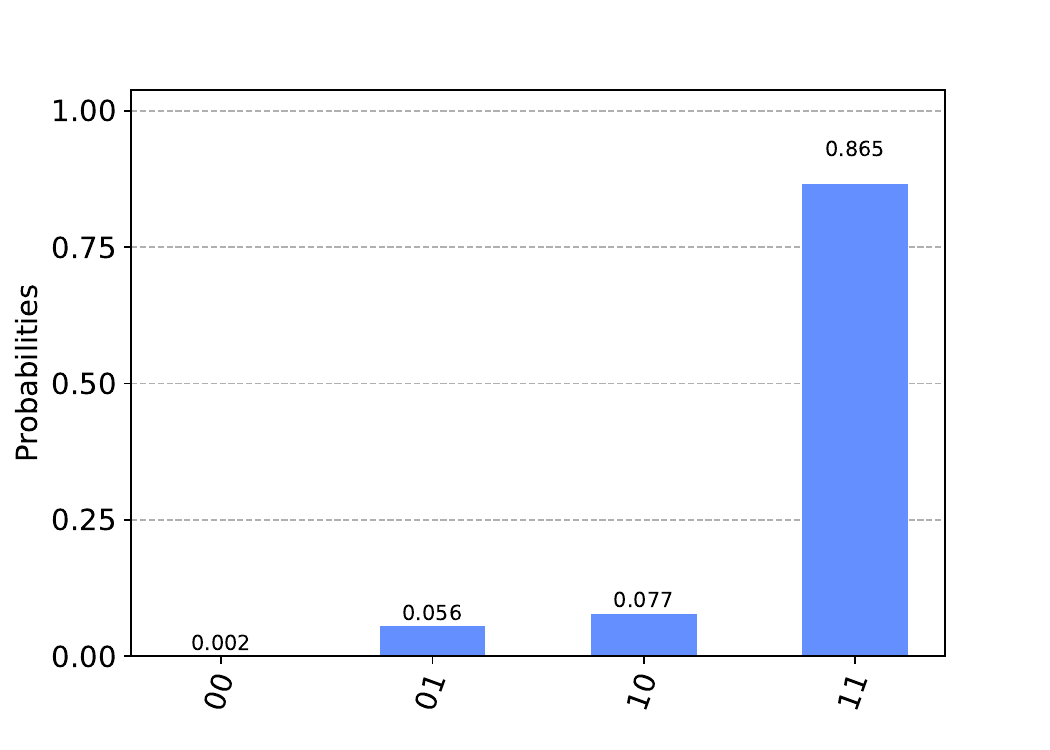}}
  \centerline{(b) Probability results}
\end{minipage}
\caption{Simulation for DRC. }
\label{dcr_simulate}
\end{figure}

\begin{figure}[!t]
\centering
\begin{minipage}{0.48\linewidth}
  \centerline{\includegraphics[width=3.5cm]{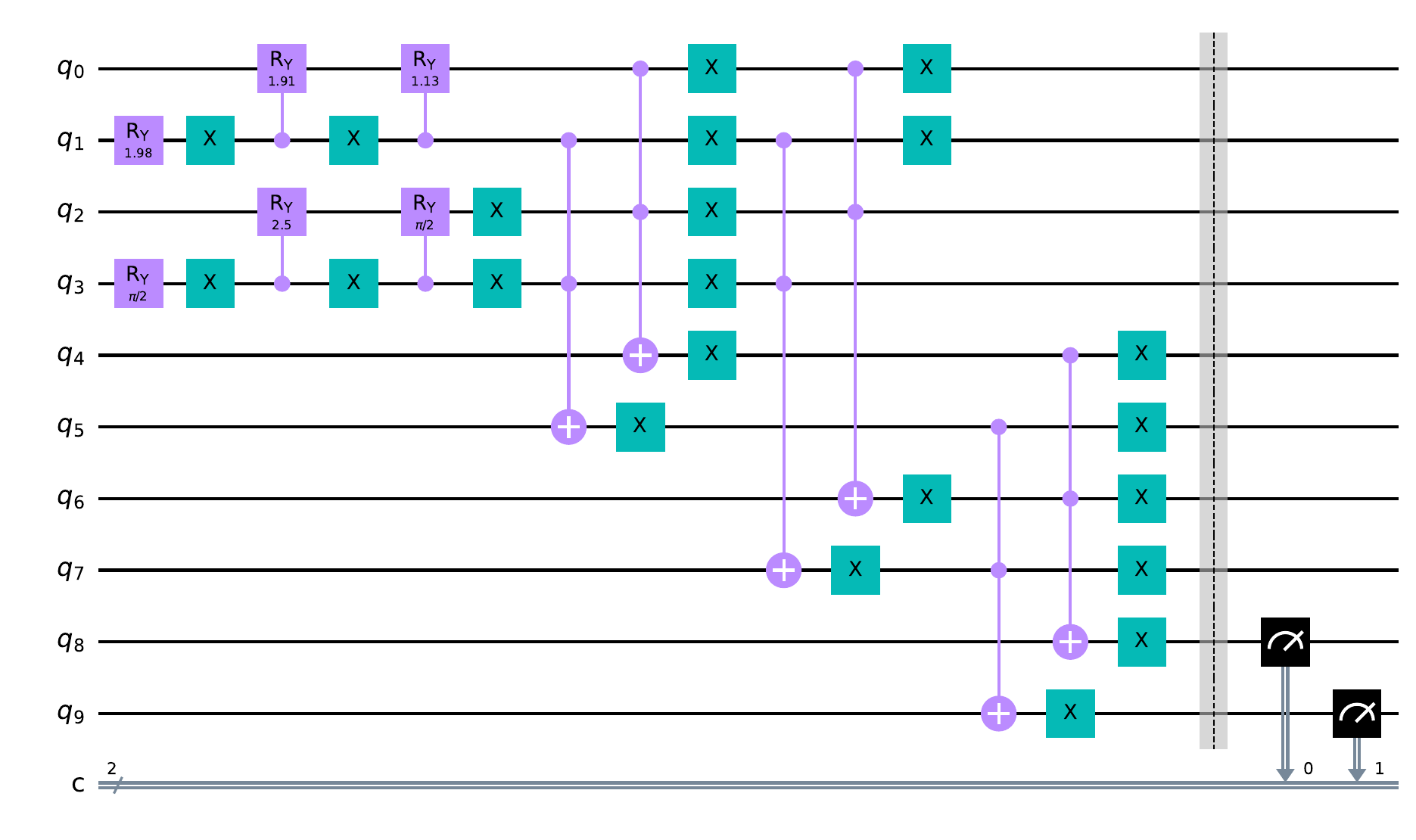}}
  \centerline{(a) Quantum circuit}
\end{minipage}
\hfill
\begin{minipage}{.48\linewidth}
  \centerline{\includegraphics[width=3.5cm]{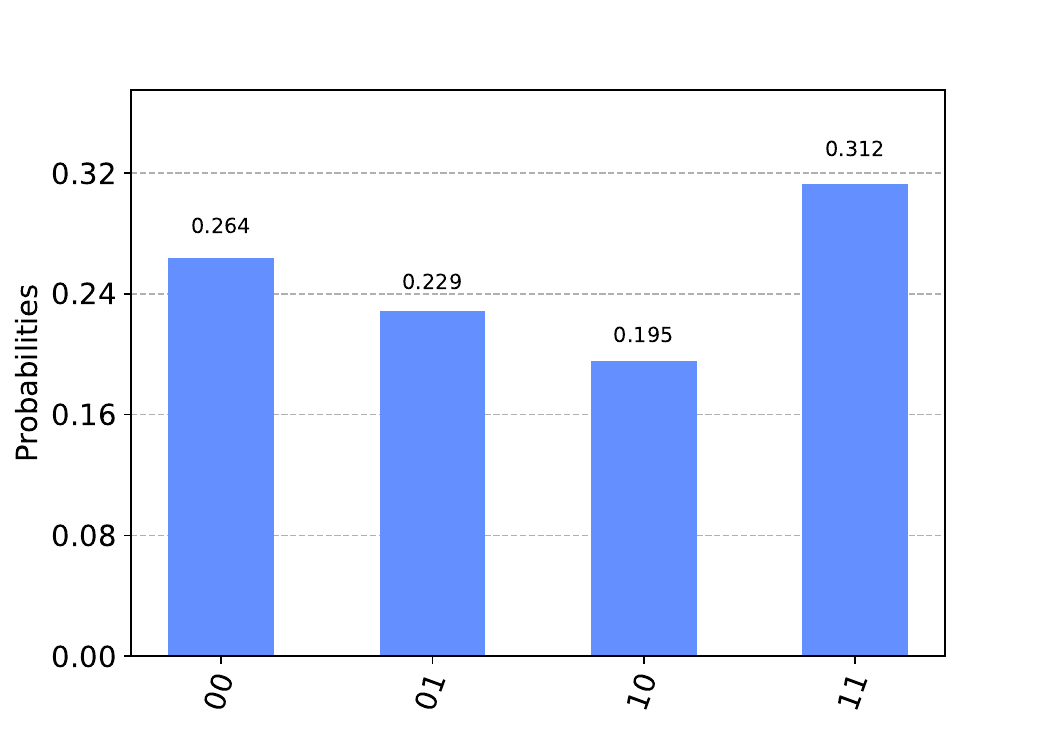}}
  \centerline{(b) Probability results}
\end{minipage}
\caption{Simulation for the exclusive disjunctive rule of combination. }
\label{exc_simulate}
\end{figure}

\begin{figure}[!t]
\centering
\begin{minipage}{0.48\linewidth}
  \centerline{\includegraphics[width=3.5cm]{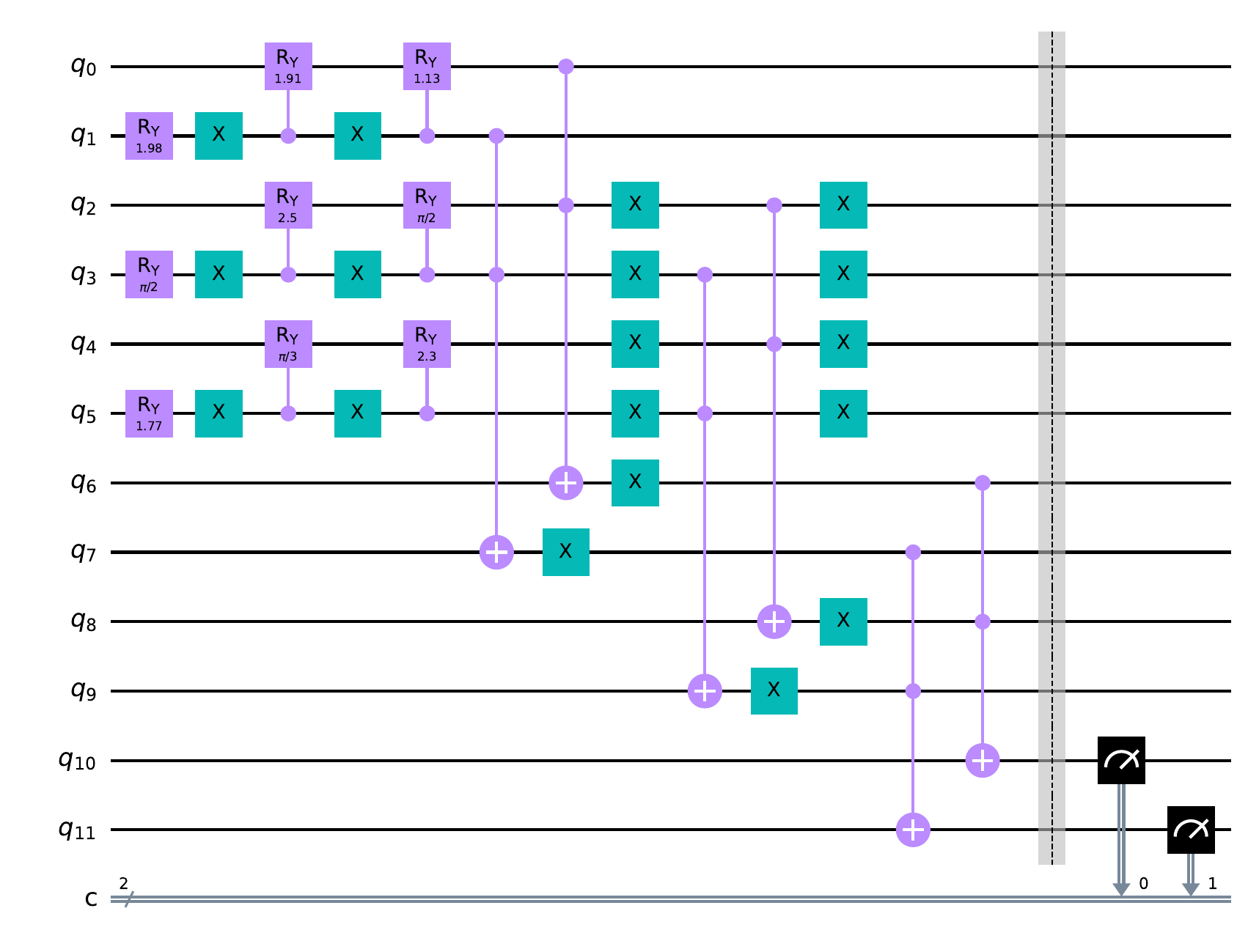}}
  \centerline{(a) Quantum circuit}
\end{minipage}
\hfill
\begin{minipage}{.48\linewidth}
  \centerline{\includegraphics[width=3.5cm]{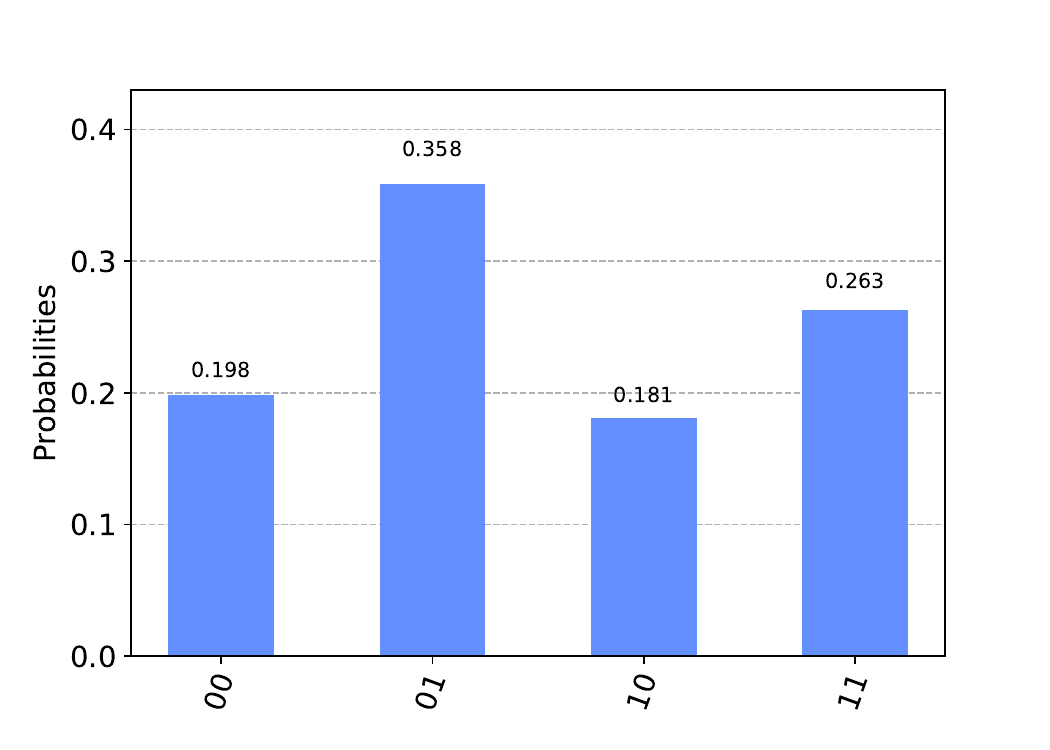}}
  \centerline{(b) Probability results}
\end{minipage}
\caption{Simulation for the customized rule of combination. }
\label{cus_simulate}
\end{figure}

Given three mass function $m_1,m_2,m_3$ under an FoD of $A$ and $B$. For each focal set $\emptyset,A,B,AB$, the mass value is set as $m_1=[0.1,0.2,0.5,0.2]$, $m_2=[0.05,0.45,0.25,0.25]$, $m_3=[0.3,0.1,0.1,0.5]$. According to the quantum circuits displayed in Fig.\ref{diag_ccr}-\ref{diag_custom}, we respectively construct the quantum circuit for the rules of combination including (1) CRC to obtain  $m_1 \circledtiny{$\cap$} m_2 \circledtiny{$\cap$} m_3$, (2) DRC to obtain $m_1 \circledtiny{$\cup$} m_2 \circledtiny{$\cup$} m_3$, (3) the exclusive disjunctive rule to obtain $m_1\circledsmall{$\underline{\cup}$}m_2$, (4) the customized rule to obtain $(\overline{m_1\circledsmall{${\cap}$}m_2})\circledsmall{${\cap}$}( m_2\circledsmall{${\cup}$}m_3)$. As our method suggests, one can directly get the combined mass value by conducting measurements (1024 shots) and calculating the probability for every ground state. The quantum circuits and the simulation result are shown in Fig.\ref{ccr_simulate}-\ref{cus_simulate}.

\begin{table}[!t]
\centering
\label{tab_simulation}
\caption{The summary of results in the simulation and the comparison with the actual mass  for four different rules of combination.}
\centering
\begin{tabular}{c|ccc|ccc}

\multicolumn{1}{c|}{\multirow{2}{*}{\makecell{Focal  \\set}}} & \multicolumn{3}{c|}{Conjunctive }                                              & \multicolumn{3}{c}{Disjunctive  }                                               \\
\multicolumn{1}{c|}{}                  & \multicolumn{1}{c|}{Simulated} & \multicolumn{1}{c|}{Actual} & Error  & \multicolumn{1}{c|}{Simulated} & \multicolumn{1}{c|}{Actual} & Error   \\ \hline
\multicolumn{1}{c|}{$\emptyset$}                 & \multicolumn{1}{c|}{0.651}     & \multicolumn{1}{c|}{0.647}  & 0.004  & \multicolumn{1}{c|}{0.002}     & \multicolumn{1}{c|}{0.0015} & 0.0005  \\ \hline
\multicolumn{1}{c|}{$A$}                 & \multicolumn{1}{c|}{0.140}     & \multicolumn{1}{c|}{0.143}  & -0.003 & \multicolumn{1}{c|}{0.056}     & \multicolumn{1}{c|}{0.0585} & -0.0025 \\ \hline
\multicolumn{1}{c|}{$B$}                 & \multicolumn{1}{c|}{0.184}     & \multicolumn{1}{c|}{0.185}  & -0.001 & \multicolumn{1}{c|}{0.077}     & \multicolumn{1}{c|}{0.0705} & 0.0065  \\ \hline
\multicolumn{1}{c|}{$AB$}                 & \multicolumn{1}{c|}{0.025}     & \multicolumn{1}{c|}{0.025}  & 0      & \multicolumn{1}{c|}{0.865}     & \multicolumn{1}{c|}{0.8695} & -0.0045 \\ \midrule[2pt]
\multirow{2}{*}{\makecell{Focal  \\set}}                       & \multicolumn{3}{c|}{Exclusive disjunctive}                                              & \multicolumn{3}{c}{Customized}                                               \\  
                                        & \multicolumn{1}{c|}{Simulated} & \multicolumn{1}{c|}{Actual} & Error  & \multicolumn{1}{c|}{Simulated} & \multicolumn{1}{c|}{Actual} & Error   \\ \hline
$\emptyset$                                       & \multicolumn{1}{c|}{0.264}     & \multicolumn{1}{c|}{0.27}   & -0.006 & \multicolumn{1}{c|}{0.198}     & \multicolumn{1}{c|}{0.207}  & -0.009  \\ \hline
$A$                                      & \multicolumn{1}{c|}{0.229}     & \multicolumn{1}{c|}{0.23}   & -0.001 & \multicolumn{1}{c|}{0.358}     & \multicolumn{1}{c|}{0.343}  & 0.015   \\ \hline
$B$                                      & \multicolumn{1}{c|}{0.195}     & \multicolumn{1}{c|}{0.19}   & 0.005  & \multicolumn{1}{c|}{0.181}     & \multicolumn{1}{c|}{0.193}  & -0.012  \\ \hline
$AB$                                      & \multicolumn{1}{c|}{0.312}     & \multicolumn{1}{c|}{0.31}   & 0.002  & \multicolumn{1}{c|}{0.263}     & \multicolumn{1}{c|}{0.257}  & 0.006   \\ 
\end{tabular}
\end{table}

To verify the feasibility of the method, in table \ref{tab_simulation} the simulated result is compared to the value of the actual combined mass functions, which are directly calculated by the set-theoretic definitions in Eq.(\ref{ccr_def_n}-\ref{customized}). From the table data, the results obtained by measurement match the actual values. Certain errors encountered in the simulation are reasonable due to the limited number of shots (here is 1024) when measuring. One can increase the number of shots to extract more accurate amplitude information for qubits thus reducing the error.

\section{\label{sec5}Attribute Fusion-based Evidential Classifier}
As the proposed quantum circuits structure suggests, the computation of implementing the combination rule has been exponentially accelerated on quantum circuits, where the complexity is $\mathcal{O}(n)$ under an FoD of $n$ elements. Thus, we introduced this efficient quantum circuit structure into the procedure of the classical classification method \cite{Hu2023attribute}  to construct an evidential classifier on quantum circuits.

Assume that the data set be divided into $n$ classes denoting $\theta^i,i=1,\dots,n$, which compose a FoD: $\Theta=\{ \theta^1,\dots,\theta^n \}$. All the data have $m$ attributes noted as $j=1,\dots,m$. Take the classic Iris data set as an example. The classes are Setosa ($\theta^1$), Versicolour ($\theta^2$), and Virginica ($\theta^3$) and the attributes are Sepal Length ($j=1$), Sepal Width ($j=2$), Petal Length ($j=3$), Petal Width ($j=4$), respectively. The evidential classifier consists of several components, which are discussed in detail below.

\subsection{Procedure}
\subsubsection{Establish Gaussian Mixture Model (GMM)}
GMM is a model of the sum of multiple Gaussian distributions, which is rich enough to fit any non-Gaussian probability distribution function for a variable \cite{Alspach1972}. GMM's function expression is given by:
\begin{equation}\label{GMM}
\begin{aligned}
        f(x)&=\sum_{k=1}^N w_k N(x|\mu_k,\sigma_k)\\
        &=\sum_{k=1}^N w_k \frac{1}{\sqrt{2\pi \sigma_k^2}}\exp \left( -\frac{(x-\mu_k)^2}{2\sigma_k^2}\right),
\end{aligned}
\end{equation}
where $f(x)$ contains $N$ Gaussian components. $\mu_k$ and $\sigma_k$ are the mean and variance of each component. $w_k$ represents the weight associated with each component, satisfying $\sum_{k=1}^N w_k=1$.

A portion of the whole dataset is extracted as the training set to derive GMM's function $f_j^i(x)$ for each attribute $j$ and class $\theta^i$. Thus there are $m\times n$ independent GMM to be established. To determine the parameters $w_k, \mu_k,\sigma_k$ in GMM, we utilize Expectation-maximization (EM) algorithm, which is summarized in the following steps. Suppose the input data is noted as $x_l$ where $l=1,\dots,L$.
\begin{enumerate}
    \item Initialization of parameters: $w_k, \mu_k,\sigma_k$, where $k=1,\dots,N$.
    \item E-step: Implement pseudo-posterior estimation. With the current model parameters, calculate the probability of considering $x_l$ to belong to $k$-th component. 
    \begin{equation}
        \gamma_{lk}=\frac{w_k N(x_l|\mu_k,\sigma_k)}{\sum_{r=1}^Nw_r N(x_l|\mu_r,\sigma_r)}.
    \end{equation}
    \item M-step: Re-estimate model parameters with the Maximum Likelihood Estimation (MLE). The updated parameters $w_k^\prime,\mu_k^\prime,\sigma_k^\prime$ can be determined by:
    \begin{equation}
    \begin{aligned}
            w_k^\prime &= \frac{1}{L} \sum_{l=1}^L \gamma_{lk};\\
        \mu_k^\prime &= \frac{\sum_{l=1}^L \gamma_{lk} x_l}{\sum_{l=1}^L \gamma_{lk}};\\
        \sigma_k^\prime &= \frac{\sum_{l=1}^L \gamma_{lk} (x_l-\mu_k)^2}{\sum_{l=1}^L \gamma_{lk}}.
    \end{aligned}
    \end{equation}
    \item Loop E-step and M-step until parameters converge.
\end{enumerate}

\begin{figure}[H]
\centering
\subfigure[Sepal Length ($j=1$)]{\includegraphics[width=.47\columnwidth]{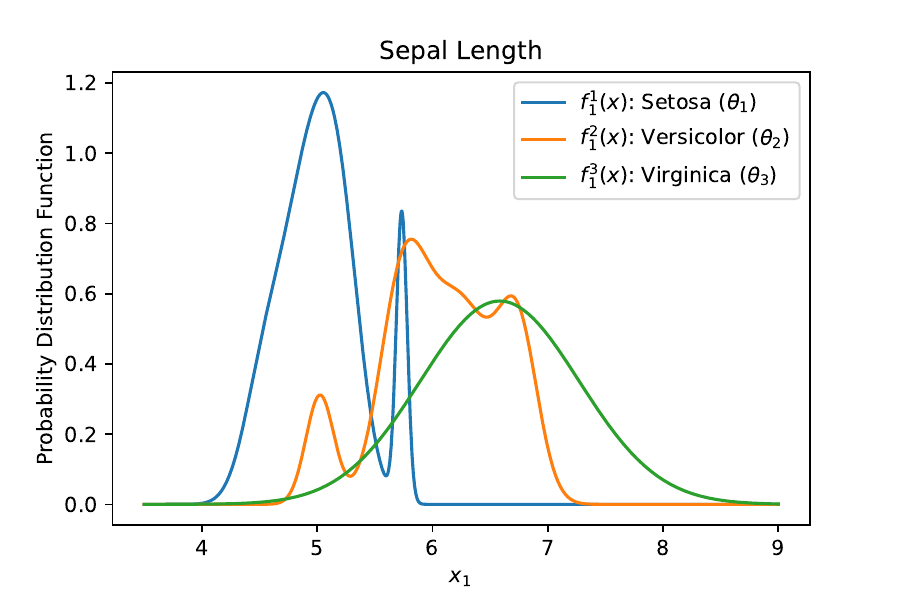}}
\hfill
\subfigure[Sepal Width ($j=2$)]{\includegraphics[width=.47\columnwidth]{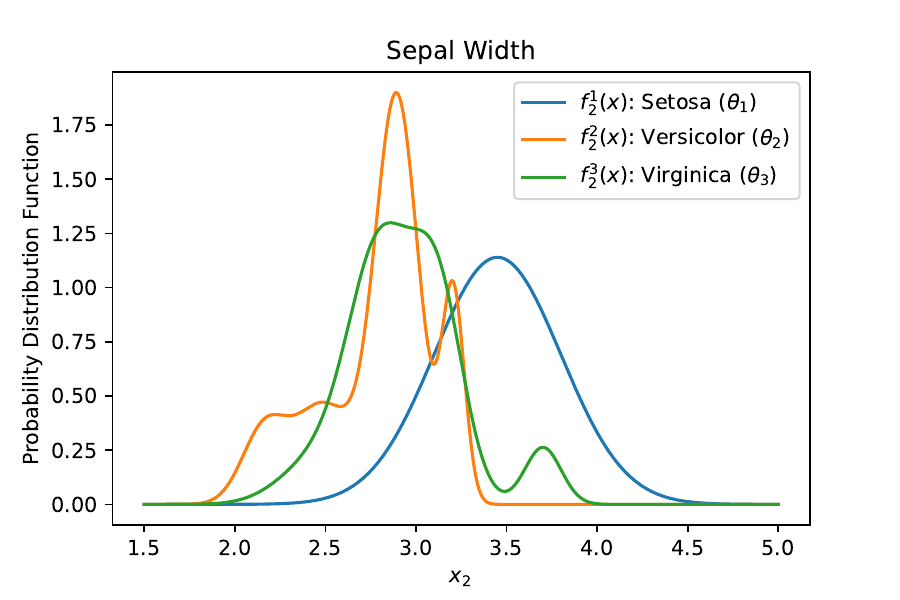}}
\\
\subfigure[Petal Length ($j=3$)]{\includegraphics[width=.47\columnwidth]{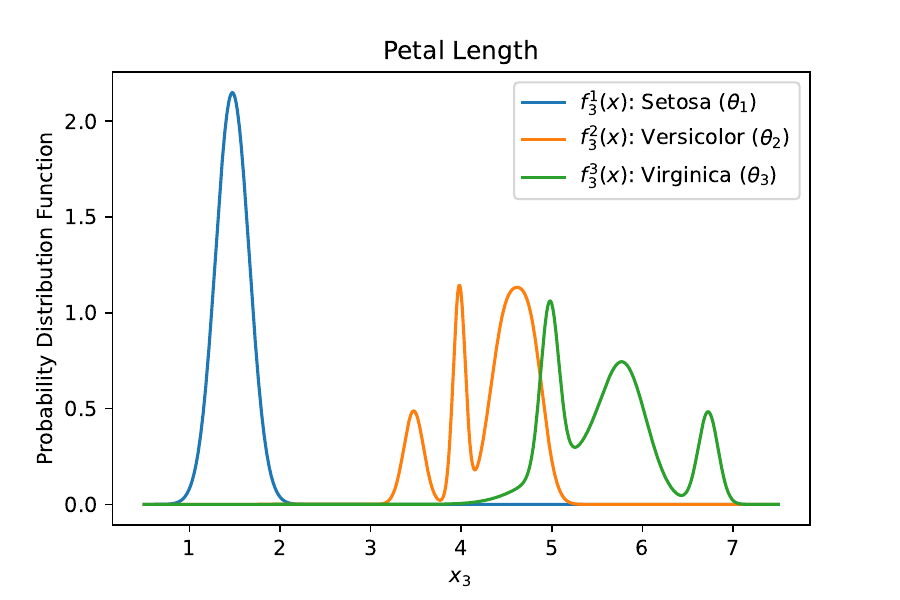}}
\hfill
\subfigure[Petal Width ($j=4$)]{\includegraphics[width=.47\columnwidth]{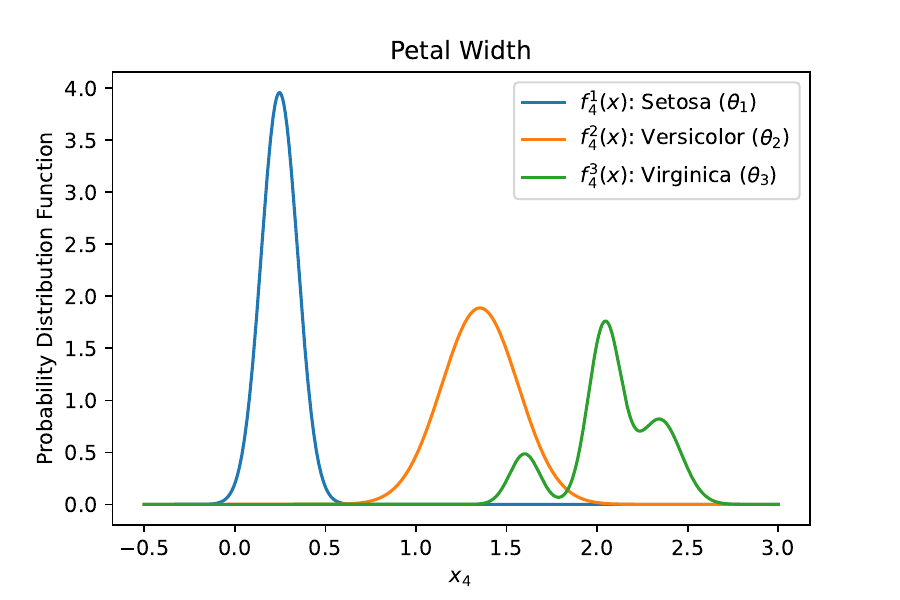}}
\caption{\label{diag_GMM} GMM for Iris data set. }
\end{figure}

For the example of the Iris data set, $3\times 4=12$ GMM should be established by EM algorithm for every attribute and class. Set the number of GMM components as $N=4$ the established GMM is shown in Fig.\ref{diag_GMM}.

\subsubsection{Generate Mass Function}
After the training stage of GMM, in the classification stage, each attribute $j$ of the input data can be evaluated by the GMM model and obtain the probability density $f_j^i(x_j)$ for each class $\theta^i$.  Then, we transform the probability density to Possibility Mass Function (PossMF) by normalization:
\begin{equation}\label{transPossMF}
\begin{aligned}
     \pi_{1j}^i=\frac{f_j^i(x_j)}{\max_k \left\{ f_j^k(x_j) \right\}},\quad\pi_{0j}^i=1-\pi_{1i}^j,
\end{aligned}
\end{equation}
where $\pi_{1j}^i$ and $\pi_{0j}^i$ represent the probability of supporting and not supporting class $\theta^j$ respectively.

\begin{figure}[H]
\centering
\includegraphics[width=5cm]{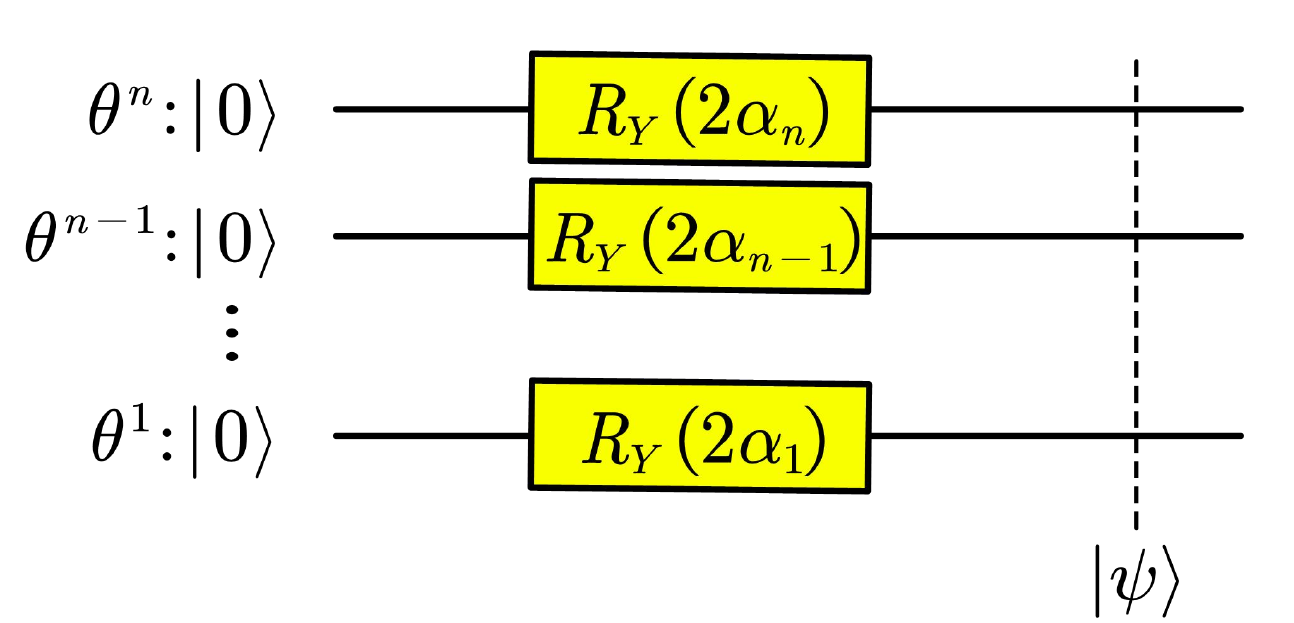}
\caption{\label{diag_simple} Simple structure quantum circuits.}
\end{figure}

Next, to handle the uncertainty in DST framework, a mass function is required to be generated according to PossMF. If the mass function is generated arbitrarily, one needs to use the tree-memory structure in Fig.\ref{diag_tree} to prepare the quantum mass function in Eq.(\ref{qmf}). However, the preparing method involves exponential computational complexity $\mathcal{O}(n2^n)$, which invalidates computational advantage in the application. To implement with lower complexity, we utilize a special method CD-BFT \cite{zhou2023cd} to generate mass function, which is given by:
\begin{equation}\label{PossMFtoM}
    m_j(\mathcal{F}(i^n\dots i^1))=\prod_{k=1}^n \pi_{i^kj}^{k},\quad j=1,\dots,m.
\end{equation}
CD-BFT is advantageous in information modeling and guarantees the combination rule consistency between possibilistic and evidential information. With the special mass structure generated by CD-BFT, one can use simpler structured quantum circuits as shown in Fig.\ref{diag_simple}.

According to the circuits,  $R_Y$ gates are only applied for $n$ times independently without entanglement introduced. And the rotation angle $\alpha_k$ can be directly determined by PossMF:
\begin{equation}\label{alpha_PossMF}
    \alpha_k=\arctan\left(\sqrt{\pi_{1j}^k / \pi_{0j}^k}\right).
\end{equation}

The step is presented in Fig.\ref{diag_test}. Take the Iris data set as an example. Suppose the input data $x=[6.3,3.3,4.7,1.6]$, where each value corresponds to an attribute. Analyze the attribute Petal Length ($j=3$), whose GMM functions are presented in Fig.\ref{diag_GMM}. Since the attribute value $x_3=4.7$, according to Fig.\ref{diag_GMM}, the probability density is $f_3^1(x_3)=0.0000$, $f_3^2(x_3)=1.1000$, $f_3^3(x_3)=0.1245$. After normalization PossMF is obtained: $\pi_{13}^1=0.0000,\pi_{03}^1=1.0000$; $\pi_{13}^2=1.0000,\pi_{03}^2=0.0000$;$\pi_{13}^2=0.1132,\pi_{03}^2=0.8868$. Finally, as Eq.(\ref{alpha_PossMF}) suggests, the angles can be determined: $\alpha_1=0$, $\alpha_2=\pi/2$, $\alpha_3=0.3431$.

\subsubsection{Evidence Combination and Decision Making}
\begin{figure}[H]
  \begin{minipage}[htbp]{0.49\columnwidth}
    \centering
    \includegraphics[width=0.98\textwidth]{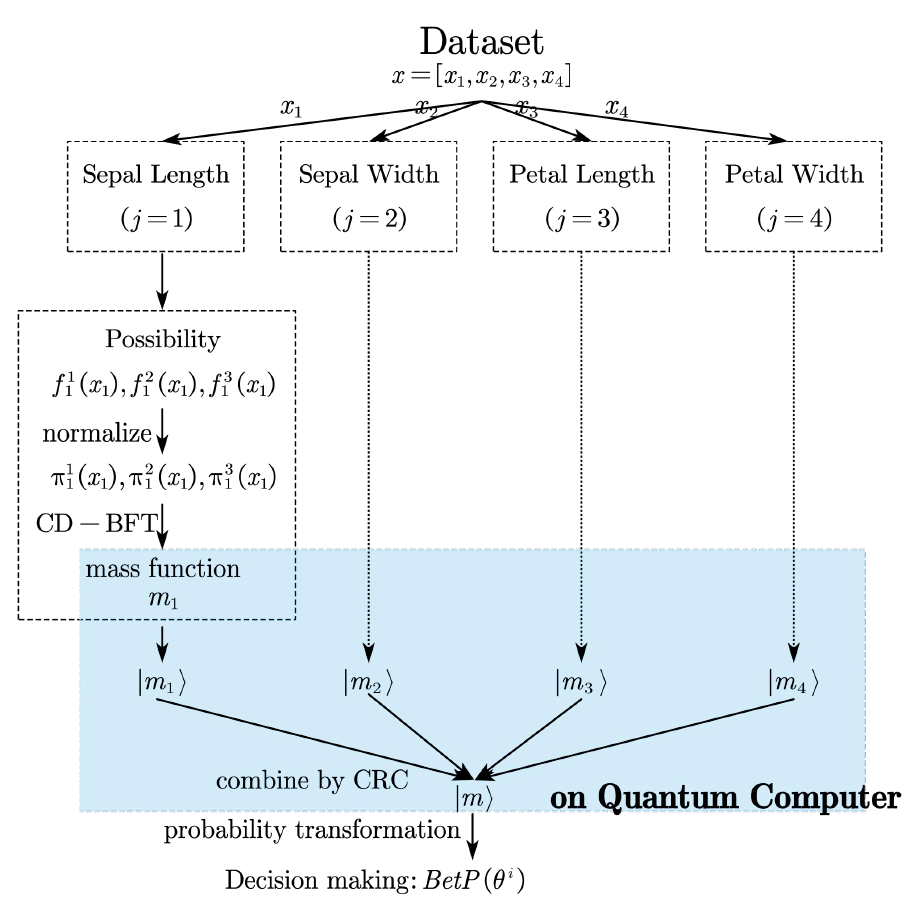}
    \caption{\label{diag_test} Flowchart for proposed attribute fusion-based  evidential classifier (use Iris dataset as an example). }
  \end{minipage}
  \begin{minipage}[htbp]{0.49\columnwidth}
    \centering
    \includegraphics[width=0.98\textwidth]{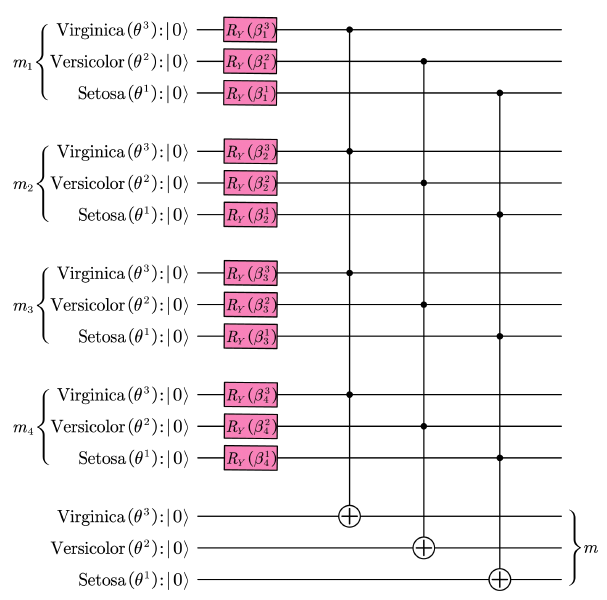}
\caption{\label{diag_whole}The quantum circuits for implementing evidence combinations involved in the evidential classifier.  (use Iris dataset as an example).}
  \end{minipage}
\end{figure}

As presented in Fig.\ref{diag_test}, the generated quantum mass functions for all attributes are fused together by CRC to achieve attribute fusion, which is given by: $m = m_1 \circledtiny{$\cap$} \dots \circledtiny{$\cap$} m_m$. Here we utilize the proposed quantum algorithm of CRC for efficient implementation. Then the combined evidence $m$ is extracted from quantum states through measurement.

To make the decision, the final step is to convert $m$ to the probability representation $BetP$, which is implemented by:
\begin{equation}\label{BetP}
    BetP(\theta^i) = \sum_{\theta_i\in F}\frac{1}{1-m(\emptyset)}\cdot \frac{m(F)}{|F|},
\end{equation}
where $|\cdot|$ denotes the number of elements contained in the focal set. Then, the final decision is $\arg\max_{\theta^i} BetP(\theta^i)$. 

For the numeral example $x = [6.3,3.3,4.7,1.6]$, after evidence combination $m$ is obtained: $m(\theta^2)=0.6725$ and $m(\theta^3\theta^2)=0.3275$. According to Eq.(\ref{BetP}), the transformed probability is: $BetP(\theta^1)=0$, $BetP(\theta^2)=0.8362$, $BetP(\theta^3)=0.1638$. Therefore, the final classification decision is that the sample belongs to the class $\theta^2$.

The entire procedure of the attribute fusion-based  evidential classifier is shown in Fig.\ref{diag_test}, where the quantum computer gets involved in the steps enclosed in the blue box with the quantum circuits as presented in Fig.\ref{diag_whole}.

\subsection{Time Complexity Analysis}
According to Fig.\ref{diag_test}, the proposed improvements over previous algorithms are the quantum implementation of classical CRC enclosed in the blue box. On quantum computers, the procedure involves Part I: generating quantum mass functions from the classical PossMF, and Part II: fusing evidence on quantum circuits. The time complexity required for previous algorithms is summarized in table \ref{tab:complexity}.

\begin{table}[H]
\caption{A comparison of time complexity.\label{tab:complexity}}
\begin{tabular}{c|cc}
Classical \cite{Hu2023attribute}  & \multicolumn{2}{c}{$\mathcal{O}(m4^n)$}                                             \\ \hline
\multirow{4}{*}{Quantum} & \multicolumn{1}{c|}{Part I}                            & Part II  \\ 
\cline{2-3}
                         & \multicolumn{1}{c|}{\multirow{2}{*}{\makecell{Tree-like Memory \cite{zhou2023bfqc}:\\ $\mathcal{O}(mn2^n)$}}} &\makecell{HHL \cite{zhou2023bfqc}: \\ $\mathcal{O}(m\log(2^{n+1}) 2^{2n} /\varepsilon)$}      \\  
                         & \multicolumn{1}{c|}{}                                  & \makecell{VQLS \cite{huangComputing2023}: \\$\mathcal{O}(m \cdot poly(2^n)\log(1/\varepsilon))$}     \\ \cline{2-3}
                         & \multicolumn{1}{c|}{Proposed: $\mathcal{O}(mn)$}                          & Proposed: $\mathcal{O}(n)$  \\ 
\end{tabular}
\end{table}

But with the proposed quantum circuit as suggested in Fig.\ref{diag_whole}, Part I is implemented by the simple structure. By acting $mn$ single qubit $R_Y$ gates, $m$ classical mass functions generated by CD-BFT can be prepared on quantum circuits. In Part II, one can use $n$ CNOT gates to obtain the combined quantum mass function. Therefore, according to table \ref{tab:complexity}, our evidential classifier achieves exponential acceleration compared to previous algorithms, reducing the total time complexity to linear: $\mathcal{O}(mn)$.

\subsection{Testing on Real Data Set}
To verify the feasibility of the proposed attribute fusion-based evidential classifier, it is tested on several real data sets \cite{Dua:2019} of different numbers of attributes and classes. Their detail is presented in table \ref{tab:dataset}.
\begin{table}[H]
\centering
\caption{\label{tab:dataset} Data Set description.}
\begin{tabular}{cccc}
Data Set & Sample: $L$ & Class: $n$ &  Attribute: $m$  \\
\hline
Iris  &150 &3 & 4\\
Haberman & 306& 2 & 3\\
Wine & 178 & 3& 13\\
Seeds & 210 & 3& 7\\
HTRU2 & 17898 & 2& 8\\
Diabetes & 768 & 2& 8\\
\end{tabular}
\end{table}

\begin{figure*}[ht]
\begin{minipage}[htbp]{0.3\linewidth}
\centering
\includegraphics[width=0.98\textwidth]{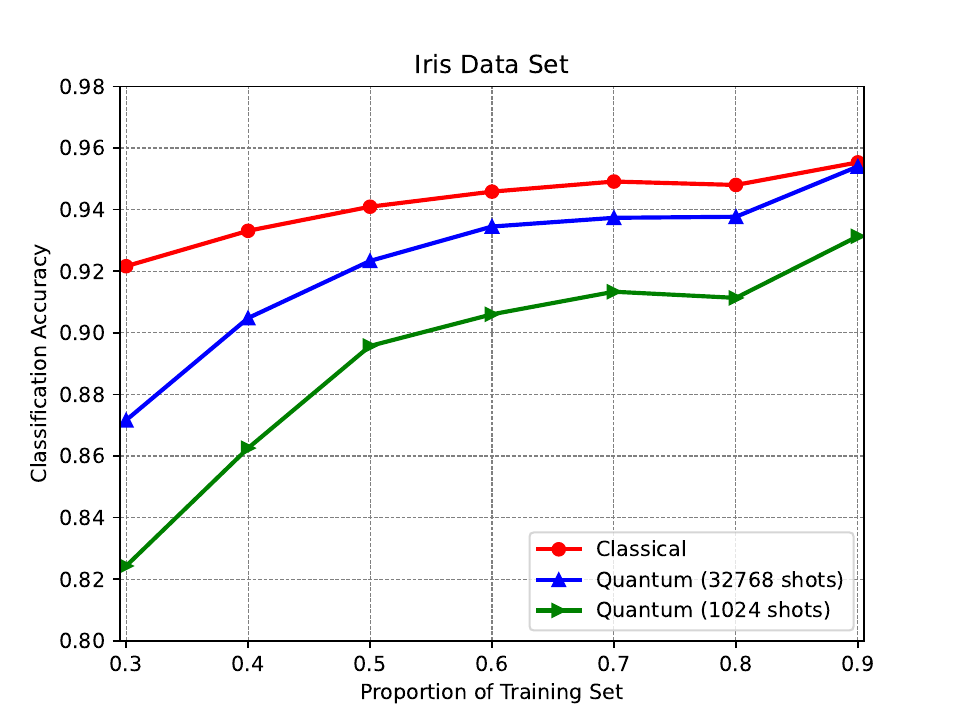}
\caption{\label{diag_iris} The classification accuracy for Iris data set.}
\end{minipage}
\hfill
\begin{minipage}[htbp]{0.3\linewidth}
\centering
\includegraphics[width=0.98\textwidth]{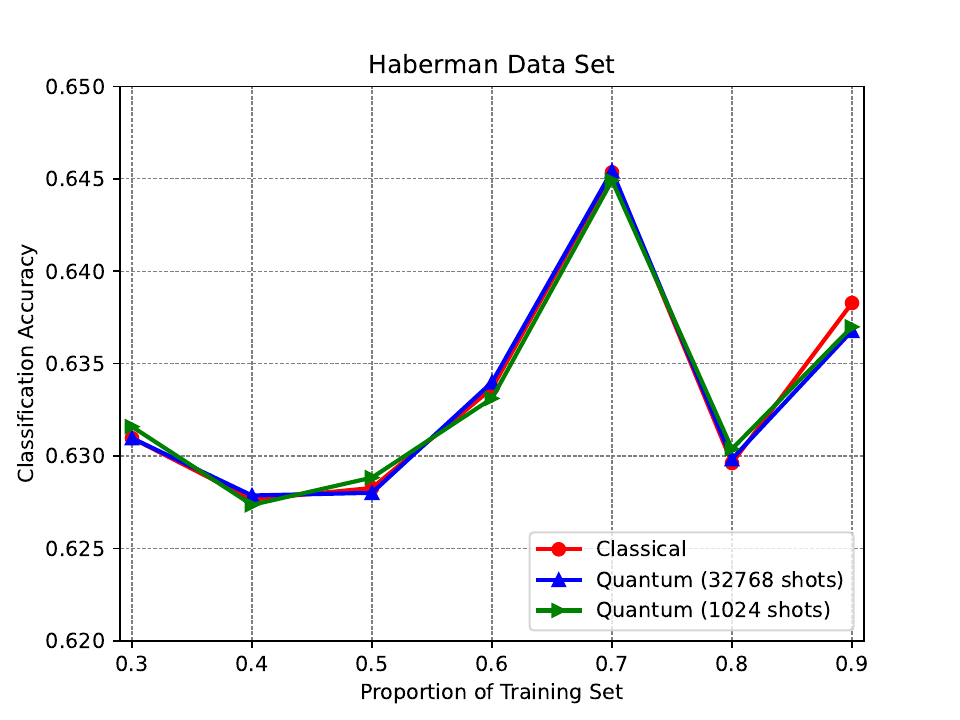}
\caption{\label{diag_hab} The classification accuracy for Haberman's Survival data set.}
\end{minipage}
\hfill
\begin{minipage}[htbp]{0.3\linewidth}
\centering
\includegraphics[width=0.98\textwidth]{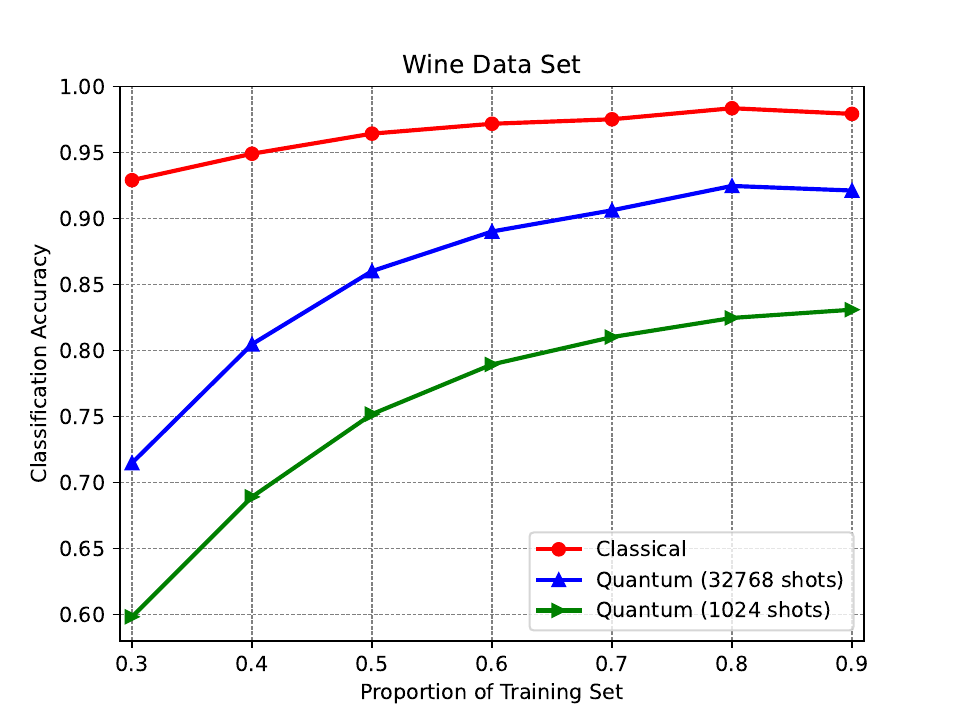}
\caption{\label{diag_wine} The classification accuracy for Wine data set.}
\end{minipage}
\\
\begin{minipage}[htbp]{0.3\linewidth}
\centering
\includegraphics[width=0.98\textwidth]{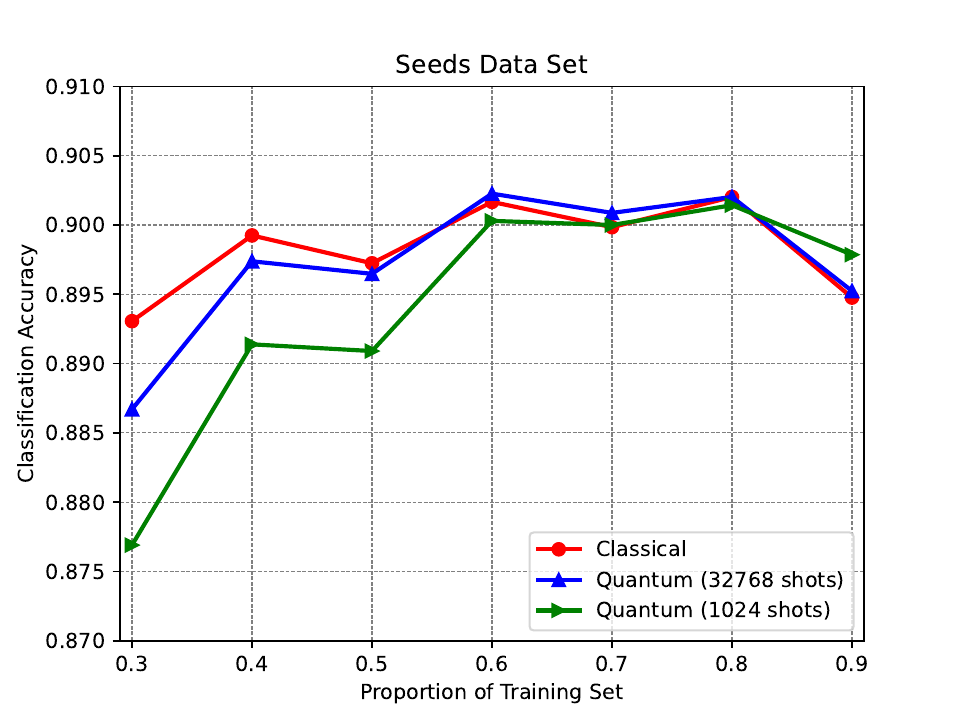}
\caption{\label{diag_seed} The classification accuracy for Seeds data set.}
\end{minipage}
\hfill
\begin{minipage}[htbp]{0.3\linewidth}
\centering
\includegraphics[width=0.98\textwidth]{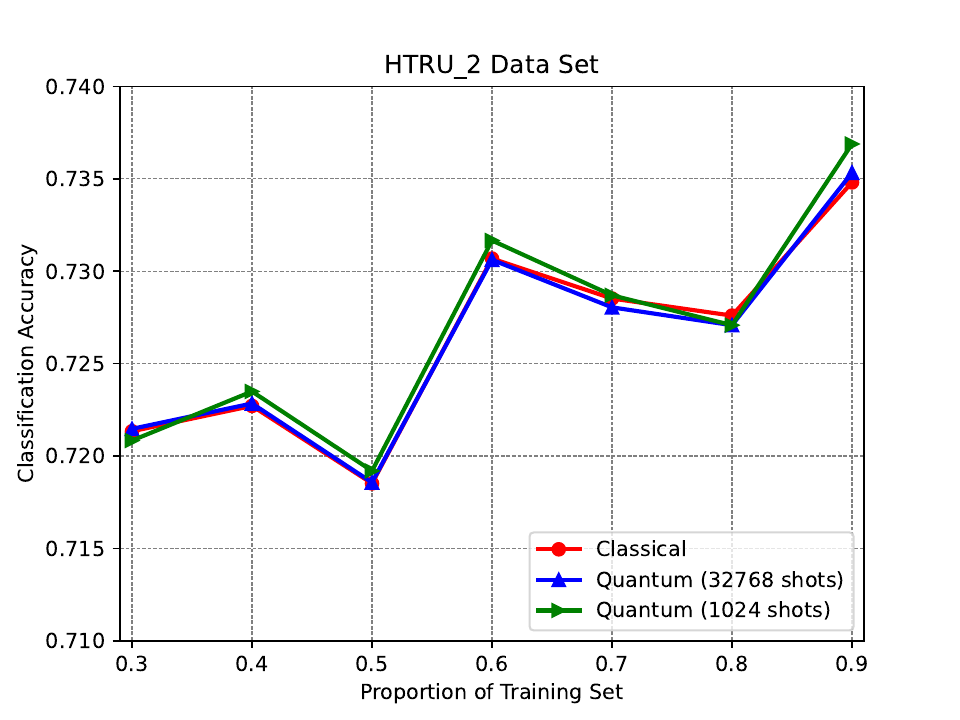}
\caption{\label{diag_htru} The classification accuracy for HTRU2 data set.}
\end{minipage}
\hfill
\begin{minipage}[htbp]{0.3\linewidth}
\centering
\includegraphics[width=0.98\textwidth]{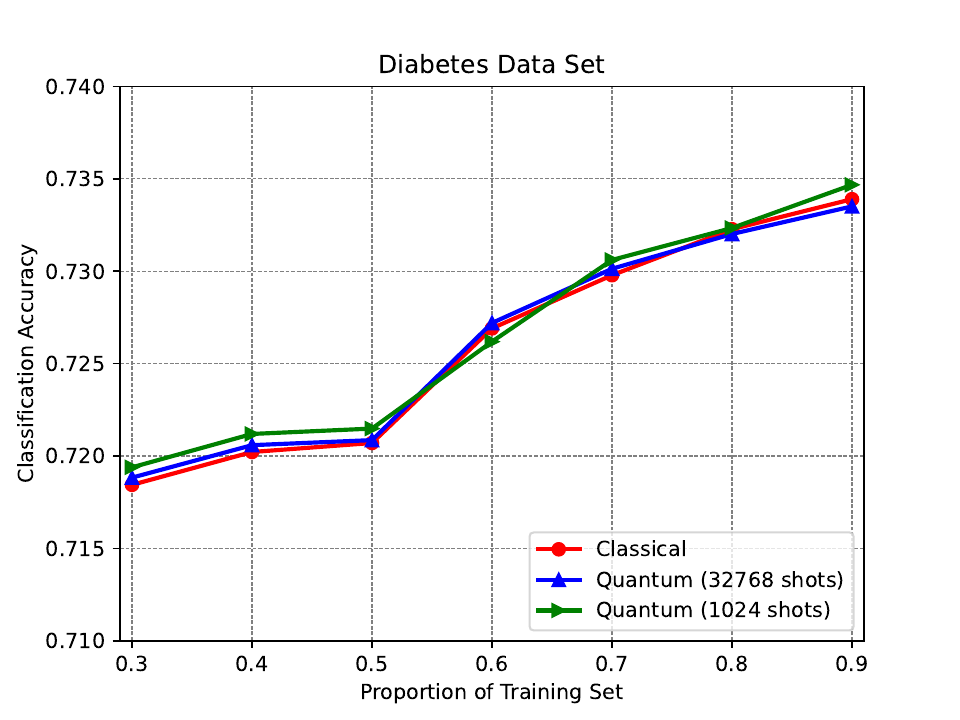}
\caption{\label{diag_diabetes} The classification accuracy for Diabetes data set.}
\end{minipage}
\end{figure*}

The proportion of data assigned to the training sets is simulated from 0.3 to 0.9. GMM component is selected as 3. Under a certain proportion, we simulate the classical attribute classifier and the proposed quantum attribute classifier by conducting 100 independent experiments. The average accuracy is shown in Fig.\ref{diag_iris}-\ref{diag_diabetes}.  Apart from the classical classifier (in red line), to demonstrate the effect of the number of shots measurement on accuracy, we select 1024 shots (in green line) and 32768 shots (in blue line) as examples. Simulation code has been uploaded to \url{https://github.com/luohaooo/DST_Classifier_on_Quantum_Circuits}.

Theoretically, the accuracy of the quantum method must be lower than that of the classical method because the proposed quantum method is based entirely on the classical model and the measurement of quantum systems brings error. And the measurement error can be eliminated by continually increasing the number of shots. However, according to the simulation results, only Iris and Wine data sets clearly demonstrate this result, especially in the case of low training proportion. As to other data sets, only a little difference in accuracy lies between classical and quantum methods, realizing ideal classification results.

\section{\label{sec6}Conclusion}
In this paper, we introduce Boolean values into the framework of DST and quantum computing and establish the correspondence between focal sets and quantum ground states. Then the set-theoretic definition of DST operations is reconsidered from a Boolean algebraic perspective. Based on that, a more flexible, efficient, and general quantum algorithm for implementing DST operations is proposed clarifying the essential link between DST and quantum computing mathematically and systematically. Simulations demonstrate the low error level of the algorithm. For the application of classification, we modify the preparation method of quantum mass functions and apply the proposed algorithm to complete the rule of combination for evidence from different attributes to obtain the attribute fusion-based evidential classifier on quantum circuits. Compared to the previous classical and quantum algorithms, the proposed classifier achieves exponential acceleration on time complexity. The final tests on real data sets verify the feasibility and accuracy of the classifier.

However, according to the proposed algorithm, the cost of reducing the time complexity is the linear increasing number of qubits required for the quantum system. Due to the current constraints on quantum resources in the current NISQ era, the practicality of this method will be limited. Therefore, reducing the size of the quantum system to design a spatially efficient method is an improvement direction. 

For future work, since the paper demonstrates a unified framework mapping DST operations to quantum circuits, more applications in the DST field \cite{gong2021evidential} can be accomplished on quantum circuits based on the proposed algorithm. From another point, the previous works focused on designing quantum circuits for accelerating DST operation. But it is time to break the limit since quantum states and operations have been given actual meaning by DST through the framework of Boolean algebra. Therefore, in the next step, we can innovate and implement the DST operations for handling uncertainty with the help of quantum computing. Further, DST brings interpretability to quantum systems, which encourages introducing DST to the current rapidly developing quantum machine learning algorithms \cite{shi2023learning,tian2023learning} and variational quantum algorithms \cite{cerezoVQA2021,huangComputing2023} to improve the performance.


{\appendix[Proof of the Zonklar Equations]
Use $\backslash${\tt{appendix}} if you have a single appendix:
Do not use $\backslash${\tt{section}} anymore after $\backslash${\tt{appendix}}, only $\backslash${\tt{section*}}.
If you have multiple appendixes use $\backslash${\tt{appendices}} then use $\backslash${\tt{section}} to start each appendix.
You must declare a $\backslash${\tt{section}} before using any $\backslash${\tt{subsection}} or using $\backslash${\tt{label}} ($\backslash${\tt{appendices}} by itself
 starts a section numbered zero.)}

{\appendices
\section{\label{appen2}Proof of Implementing CRC on Quantum Circuits}
According to Fig.\ref{diag_ccr}, the quantum system contains $(p+1)n$ qubits, including $p+1$ parts of quantum mass functions. The first $p$ parts represent $\ket{m_1},\dots,\ket{m_p}$. And the initial state of last $n$ qubits is $\ket{0}^{\otimes n}$. Thus the entire quantum state at the initial stage can be derived.
\begin{equation} \label{ccr0}
    \begin{aligned}
        \ket{\psi_0}&=\ket{m_1}\ket{m_2}\dots\ket{m_p}\ket{0}^{\otimes n}\\
&=\sum_{i_1=0}^{2^n-1}\dots \sum_{i_p=0}^{2^n-1} \prod_{r=1}^{p}\sqrt{m_r(\mathcal{F}(i_r))} \bigotimes_{r=1}^{p}\ket{i_r}_D\ket{0}^{\otimes n}.
    \end{aligned}
\end{equation}
Here $i_r (r=1,\dots,p)$ is the decimal representation of the Boolean values satisfying $i_r=(i_r^n i_r^{n-1}\dots i_r^1)_2$. Then the first CNOT gate is applied on qubits associated with $\theta^n$, flipping the target qubit to $\ket{1}$ when $i_1^n \wedge \dots i_p^n=1$. And the quantum state becomes:
\begin{equation}\label{ccr1}
    \begin{aligned}
        &\ket{\psi_1}=\\&\left\{ \sum_{i_1^n\wedge\dots\wedge i_p^n=0} \left[ 
        \sum_{i_1=0}^{2^n-1}\dots \sum_{i_p=0}^{2^n-1} \prod_{r=1}^{p}\sqrt{m_r(\mathcal{F}(i_r))} \bigotimes_{r=1}^{p}\ket{i_r}_D
        \right] \ket{0}\right.\\
        &\left.+ \sum_{i_1^n\wedge\dots\wedge i_p^n=1} \left[ \sum_{i_1=0}^{2^n-1}\dots \sum_{i_p=0}^{2^n-1} \prod_{r=1}^{p}\sqrt{m_r(\mathcal{F}(i_r))} \bigotimes_{r=1}^{p}\ket{i_r}_D \right] \ket{1}  \right\} \\ &\ket{0}^{\otimes n-1}.
    \end{aligned}
\end{equation}

Similarly, after the $k$-th CNOT operation, the quantum state $\ket{\psi_k}$ can be deduced.
\begin{equation}\label{ccrk}
    \begin{aligned}
        &\ket{\psi_k}=\\&\sum_{j^n=0}^1 \dots \sum_{j^{n-k+1}=0}^1 \left\{ \sum_{i_1^n\wedge\dots\wedge i_p^n=j^n} \dots \sum_{i_1^{n-k+1}\wedge\dots\wedge i_p^{n-k+1}=j^{n-k+1}} \right. \\
        &\left. \left[ \sum_{i_1=0}^{2^n-1}\dots \sum_{i_p=0}^{2^n-1} \prod_{r=1}^{p}\sqrt{m_r(\mathcal{F}(i_r))} \bigotimes_{r=1}^{p}\ket{i_r}_D \right]\ket{j^n\dots j^{n-k+1}} \right\}\\&\ket{0}^{\otimes n-k}.
    \end{aligned}
\end{equation}

Therefore, according to Eq.(\ref{ccrk}), when $k=n$ the final quantum state is obtained.

    \begin{equation}\label{ccrn}
    \begin{aligned}
        \ket{\psi_n}&=\sum_{j^n=0}^1 \dots \sum_{j^{1}=0}^1 \Bigg\{ \sum_{i_1^n\wedge\dots\wedge i_p^n=j^n} \dots \sum_{i_1^{1}\wedge\dots\wedge i_p^{1}=j^{1}} \Bigg[ \sum_{i_1=0}^{2^n-1}\dots \sum_{i_p=0}^{2^n-1} \\&\prod_{r=1}^{p}\sqrt{m_r(\mathcal{F}(i_r))} \bigotimes_{r=1}^{p}\ket{i_r}_D\Bigg] \ket{j^n\dots j^{1}} \Bigg\} \\
        &=\sum_{j^n=0}^1 \dots \sum_{j^{1}=0}^1 \Bigg\{\sum_{i_1^n\wedge\dots\wedge i_p^n=j^n} \dots \sum_{i_1^{1}\wedge\dots\wedge i_p^{1}=j^{1}} \\&\prod_{r=1}^{p}\sqrt{m_r(\mathcal{F}(i_r^n i_r^{n-1} \dots i_r^1))} \bigotimes_{r=1}^{p}\ket{i_r^n i_r^{n-1} \dots i_r^1}\ket{j^n\dots j^{1}}\Bigg\}.
    \end{aligned}
\end{equation}

\section{\label{appen3}Proof of Implementing DRC on Quantum Circuits}
According to Fig.\ref{diag_dcr}, in step 1, compared to the conjunctive rule, the method only takes the reverse of the input Boolean value. Therefore, according to Eq.(\ref{ccr_psin}), the quantum state $\ket{\psi_1}$ after step 1 can be derived.
\begin{equation}\label{dcr1}
\begin{aligned}
        \ket{\psi_1}&=\sum_{j^n=0}^1 \dots \sum_{j^{1}=0}^1 \Bigg\{\sum_{\neg i_1^n\wedge\dots\wedge \neg i_p^n=j^n} \dots \sum_{\neg i_1^{1}\wedge\dots\wedge \neg i_p^{1}=j^{1}} \\ &\prod_{r=1}^{p}\sqrt{m_r(\mathcal{F}(i_r^n  \dots i_r^1))} \bigotimes_{r=1}^{p}\ket{i_r^n  \dots i_r^1}\ket{j^n\dots j^{1}}\Bigg\}.
\end{aligned}
\end{equation}

In step 2, $X$ gates are implemented on the last $n$ qubits, flipping the state $\ket{j^n\dots j^1}$. The quantum state $\ket{\psi_2}$ can be expressed and processed by Boolean algebra as follows.

\begin{equation}\label{dcr2}
    \begin{aligned}
        \ket{\psi_2}&=\sum_{j^n=0}^1 \dots \sum_{j^{1}=0}^1 \Bigg\{\sum_{\neg i_1^n\wedge\dots\wedge \neg i_p^n=j^n} \dots \sum_{\neg i_1^{1}\wedge\dots\wedge \neg i_p^{1}=j^{1}} \\&\quad\quad\prod_{r=1}^{p}\sqrt{m_r(\mathcal{F}(i_r^n  \dots i_r^1))} \bigotimes_{r=1}^{p}\ket{i_r^n  \dots i_r^1}\ket{\neg j^n\dots \neg j^{1}}\Bigg\}\\
        &=\sum_{j^n=0}^1 \dots \sum_{j^{1}=0}^1 \Bigg\{\sum_{\neg i_1^n\wedge\dots\wedge \neg i_p^n=\neg j^n} \dots \sum_{\neg i_1^{1}\wedge\dots\wedge \neg i_p^{1}=\neg j^{1}} \\ &\quad \quad\prod_{r=1}^{p}\sqrt{m_r(\mathcal{F}(i_r^n  \dots i_r^1))} \bigotimes_{r=1}^{p}\ket{i_r^n  \dots i_r^1}\ket{j^n\dots j^{1}}\Bigg\}\\
        &=\sum_{j^n=0}^1 \dots \sum_{j^{1}=0}^1 \Bigg\{\sum_{ i_1^n\vee\dots\vee  i_p^n= j^n} \dots \sum_{ i_1^{1}\vee\dots\vee  i_p^{1}= j^{1}} \\&\quad \quad\prod_{r=1}^{p}\sqrt{m_r(\mathcal{F}(i_r^n  \dots i_r^1))} \bigotimes_{r=1}^{p}\ket{i_r^n  \dots i_r^1}\ket{j^n\dots j^{1}}\Bigg\}.
    \end{aligned}
\end{equation}
}

\bibliographystyle{IEEEtran}
\bibliography{IEEEabrv,bibtex}

\end{document}